\documentclass[a4paper,fleqn,11pt]{article}
\pdfoutput=1
\usepackage{amsmath}
\usepackage{amssymb}
\usepackage{array}
\usepackage{calc}
\usepackage{longtable}
\usepackage{multirow}
\usepackage{slashed}
\usepackage{booktabs}
\usepackage{pstricks}
\usepackage{graphicx}
\usepackage{xspace}
\usepackage{units}
\usepackage{cite}
\usepackage{textcomp}

\numberwithin{equation}{section}
\usepackage[pdfborder={0 0 0}]{hyperref}
\usepackage[format=hang,labelfont=bf,hypcap=true]{caption}
\usepackage{subcaption}
\usepackage{sectsty}
\usepackage{enumitem}
\allsectionsfont{\sffamily}
\subsubsectionfont{\mdseries\itshape\large}
\makeatletter
\DeclareRobustCommand*{\bfseries}{%
  \not@math@alphabet\bfseries\mathbf
  \fontseries\bfdefault\selectfont
  \boldmath
}
\makeatother
\let\spreprint\empty
\newcommand{\preprint}[1]{\def\spreprint{\protect#1}}
\let\sinstitute\empty
\newcommand{\institute}[1]{\def\sinstitute{\protect#1}}
\makeatletter
\renewcommand{\maketitle}{\begingroup
  \null\thispagestyle{empty}%
    \ifx\spreprint\empty
      \vskip 5ex
    \else
      \flushright\large\spreprint\vskip 10ex
    \fi
    \vskip 5ex
    \flushleft
      {\sffamily\bfseries\huge\@title}\vskip 6ex
      \@author\vskip 2ex
      \ifx\sinstitute\empty
      \else
        {\small\sinstitute}
      \fi
    \vskip 5ex
  \endgroup
}
\makeatother
\renewenvironment{abstract}{\begin{center}
  {\large\sffamily\bfseries Abstract: }
  \begin{minipage}[t]{0.75\textwidth}
}{\end{minipage}\end{center}\vskip 10ex}

\numberwithin{equation}{section}
\allowdisplaybreaks[2]
\newcommand{\MINLO}{MiN\protect\scalebox{0.8}{LO}\xspace}

\newcommand{\POWHEG}{P\protect\scalebox{0.8}{OWHEG}\xspace}

\newcommand{\Herwig}{H\protect\scalebox{0.8}{ERWIG}\xspace}

\newcommand{\MadGraphFive}{M\protect\scalebox{0.8}{AD}G\protect\scalebox{0.8}{RAPH}5\xspace}

\newcommand{\FeynCalc}{F\protect\scalebox{0.8}{EYN}C\protect\scalebox{0.8}{ALC}\xspace}
\newcommand{\FormCalc}{F\protect\scalebox{0.8}{ORM}C\protect\scalebox{0.8}{ALC}\xspace}
\newcommand{\PackageX}{\textit{Package}-\texttt{X}\xspace}
\newcommand{\Collier}{C\protect\scalebox{0.8}{OLLIER}\xspace}
\newcommand{\QCDLoop}{\texttt{QCDLoop}\xspace}

\newcommand{\Rivet}{R\protect\scalebox{0.8}{IVET}\xspace}

\newcommand{\GoSam}{G\protect\scalebox{0.8}{O}S\protect\scalebox{0.8}{AM}\xspace}

\newcommand{\Recola}{R\protect\scalebox{0.8}{ECOLA}\xspace}
\newcommand{\OpenLoops}{O\protect\scalebox{0.8}{PEN}L\protect\scalebox{0.8}{OOPS}\xspace}

\newcommand{\FEWZ}{F\protect\scalebox{0.8}{EWZ}\xspace}
\newcommand{\Horace}{H\protect\scalebox{0.8}{ORACE}\xspace}
\newcommand{\Photos}{P\protect\scalebox{0.8}{HOTOS}\xspace}

\newcommand{\Rady}{R\protect\scalebox{0.8}{ADY}\xspace}
\newcommand{\Sanc}{S\protect\scalebox{0.8}{ANC}\xspace}
\newcommand{\Winhac}{W\protect\scalebox{0.8}{INHAC}\xspace}
\newcommand{\WZGRAD}{W\protect\scalebox{0.8}{ZGRAD}\xspace}

\newcommand{\DYNNLOPS}{D\protect\scalebox{0.8}{YNNLOPS}\xspace}

\newcommand{\Sophty}{S\protect\scalebox{0.8}{OPHTY}\xspace}
\newcommand{\Sherpa}{S\protect\scalebox{0.8}{HERPA}\xspace}
\newcommand{\Comix}{C\protect\scalebox{0.8}{OMIX}\xspace}

\newcommand{\Amegic}{A\protect\scalebox{0.8}{MEGIC}\xspace}

\newcommand{\Photons}{P\protect\scalebox{0.8}{HOTONS}\xspace}

\newcommand{\LHC}{LHC\xspace}

\long\def\symbolfootnote[#1]#2{\begingroup%
\def\thefootnote{\fnsymbol{footnote}}\footnote[#1]{#2}\endgroup}
\newcommand{\EqRef}[1]{Eq.~\eqref{#1}}

\newcommand{\mc}[1]{\mathcal{#1}}
\newcommand{\mr}[1]{\mathrm{#1}}

\newcommand{\bea}{\begin{eqnarray}}
\newcommand{\eea}{\end{eqnarray}}
\newcommand{\bi}{\begin{itemize}}
\newcommand{\ei}{\end{itemize}}

\newcommand{\GeV}{\,\mathrm{GeV}}
\newcommand{\MeV}{\,\mathrm{MeV}}

\newlist{myitemize}{itemize}{3}
\setlist[myitemize]{leftmargin=14em}

\newcommand{\figscale}{0.5}

\hypersetup{
  pdfauthor={Frank Krauss, Robin Linte, Marek Schoenherr},
  pdftitle={NLO EW and NNLO QED corrected soft-photon resummation}
}
\preprint{IPPP--18--88\\CERN-TH-2018-212\\MCnet-18-26}

\title{Accurate simulation of $W$, $Z$, and Higgs boson decays in \Sherpa}

\author{Frank Krauss$^1$, Jonas M. Lindert$^1$, Robin Linten$^1$, Marek Sch{\"o}nherr$^2$}
\institute{
$^1$Institute for Particle Physics Phenomenology, University of Durham, Durham,~DH1~3LE, UK\\
$^2$Theoretical Physics Department, CERN, 1211 Geneva 23, Switzerland}
\begin{document}
\vspace*{10mm}
\maketitle
\vspace*{20mm}
\begin{abstract}
  We discuss the inclusion of next-to--\-next-to leading order electromagnetic and
  of next-to leading order electroweak corrections to the leptonic decays of weak
  gauge and Higgs bosons in the SHERPA event generator. To this end, we modify the Yennie-%
  Frautschi-Suura scheme for the resummation of soft photon corrections and its
  systematic improvement with fixed-order calculations, to also include the effect of
  virtual corrections due to the exchange of weak gauge bosons. We detail relevant
  technical aspects of our implementation and present numerical results  for
  observables relevant for high-precision Drell-Yan and Higgs boson production 
  and decay simulations at the LHC.
\end{abstract}
\newpage
\tableofcontents
\section{Introduction}
\label{sec:intro}

The experiments at the LHC are stress-testing the Standard Model (SM)
of particle physics at unprecedented levels of precision. In
particular, leptonic standard-candle signatures like charged- and
neutral-current Drell-Yan production offer large cross sections
together with very small experimental uncertainties, often at or even
below the percent level.  This allows to extract fundamental
parameters in the electroweak (EW) sector of the SM at levels of
precision surpassing the LEP heritage. Measurements of the $W$-boson
mass, a key EW precision observable, are already reaching the 20 MeV
level~\cite{Aaboud:2017svj} based on 7 TeV data alone, with theory
uncertainties being one of the leading systematics. Another example
for the impressive achievements on the experimental side, challenging
currently available theoretical precision, is the recent measurement
of the triple differential cross section in neutral current Drell-Yan
production based on 8 TeV data \cite{Aaboud:2017ffb}, the first of its
kind at a hadron collider.
Furthermore, precision measurements of the $Z$ transverse momentum
spectrum~\cite{Khachatryan:2015oaa,Aad:2015auj} have been used to
constrain parton distribution functions
(PDFs)~\cite{Boughezal:2017nla}.
In order to fully harness available and future experimental datasets
excellent theoretical control of various very subtle effects of
higher-order QCD and EW origin is required.  For recent reviews and
studies on these issues, see
e.g.\ \cite{Badger:2016bpw,Alioli:2016fum,CarloniCalame:2016ouw}. With
this paper we contribute to this effort by investigating higher-order
QED/EW effects in the modelling of soft-photon radiation off
vector-boson decays.

The demand for (sub-)percent precision in Drell-Yan production has led
to formidable achievements in the theoretical description of
corresponding collider observables, often pushing boundaries of
technical limitations. The pioneering next-to-next-to-leading (NNLO)
QCD corrections for differential Drell-Yan
production~\cite{Anastasiou:2003ds,Anastasiou:2003yy,Melnikov:2006kv}
are available as public computer
codes~\cite{Catani:2009sm,Gavin:2010az,Grazzini:2017mhc} and have
recently been matched to QCD parton showers, using the
$\mr{UN}^2\mr{LOPS}$ framework within \Sherpa~\cite{Hoeche:2014aia},
and via a reweighting of a \MINLO improved calculation in \DYNNLOPS
\cite{Karlberg:2014qua}. Since recently also NNLO corrections to
Drell-Yan production at finite transverse momentum are
available~\cite{Ridder:2015dxa,Ridder:2016nkl,Gehrmann-DeRidder:2016jns,Gehrmann-DeRidder:2017mvr,Boughezal:2015ded,Boughezal:2016isb,Boughezal:2015dva,Boughezal:2016dtm}.
Higher-order EW corrections at the NLO level for inclusive Drell-Yan
production are available for quite some
time~\cite{Wackeroth:1996hz,Baur:1997wa} and are implemented in a large
number of public codes, including \WZGRAD
\cite{Baur:1998kt,Baur:2001ze,Baur:2004ig}, \Sanc
\cite{Andonov:2004hi}, \Rady \cite{Dittmaier:2001ay,Dittmaier:2009cr}
and \FEWZ \cite{Li:2012wna}.  At finite transverse momentum they have
been calculated
in~\cite{Denner:2009gj,Denner:2011vu,Denner:2012ts,Kallweit:2015dum}.
The combination of higher-order QCD and EW effects is available within
the \POWHEG framework
\cite{Alioli:2008gx,Bernaciak:2012hj,Barze:2012tt,Barze:2013fru,CarloniCalame:2016ouw}
matched to parton-showers, and also in \cite{Muck:2016pko}.  Efforts to calculate the fixed-order NNLO 
mixed QCD and EW corrections explicitly are
underway~\cite{Bonciani:2016wya,Bonciani:2016ypc,vonManteuffel:2017myy,deFlorian:2018wcj}.
Their effect has been studied in the pole approximation
\cite{Dittmaier:2014qza,Dittmaier:2015rxo}.

At the desired level of precision QED effects impacting in
particular the leptonic final state have to be considered and
understood in detail. In this case, soft and collinear photon radiation provides
the major contributions. These can be resummed to all orders, and also
improved order by order in perturbation theory. Implementations of
such calculations have been performed via a QED parton shower matching
in \Horace \cite{CarloniCalame:2003ux} and in the \POWHEG framework,
in the structure function approach in \Rady, and through a YFS-type
exponentiation for particle decays in \Photos \cite{Barberio:1993qi},
\Winhac \cite{Placzek:2003zg}, the \Herwig module \Sophty
\cite{Hamilton:2006xz} and the \Sherpa module \Photons
\cite{Schonherr:2008av}.
In this paper, we present an extension of the \Sherpa module \Photons,
which provides a 
simulation of QED radiation in (uncoloured) particle
decays. \Photons implements the approach of Yennie, Frautschi and
Suura (YFS) \cite{Yennie:1961ad} for the calculation of higher order
QED corrections. In the YFS approach, leading soft logarithms, which
are largely independent of the actual hard process involved, are
resummed to all orders. Beyond this, the method also allows for the
systematic improvement of the description through the inclusion of
full fixed-order matrix elements.
The present implementation allows for the inclusion of a collinear
approximation to the real matrix element using dipole splitting
kernels~\cite{Dittmaier:1999mb}. Furthermore, for several relevant
processes, including the decays of electroweak bosons, $\tau$ decays
as well as generic decays of uncharged scalars, fermionic and vector
hadrons, the full real and virtual NLO QED matrix elements are
included. This module has also been used for the description of
electroweak corrections in the semileptonic decays of $B$ mesons
\cite{Bernlochner:2010fc}.  The aim of this publication is to further
enhance the level of precision in the case of the decay of electroweak
gauge- and Higgs-bosons by implementing the full one-loop EW
corrections, as well as NNLO QED corrections in the case of $Z$- and
Higgs-decays.
The electroweak virtual corrections to particle decays are known for
a long time~\cite{Denner:1991kt,Bardin:1999ak} and our implementation
will be based on these results. In the case of $Z$-boson decays, the
double virtual corrections in the limit of small lepton masses are
known for about 30 years \cite{Berends:1987ab}, which we will rely on.

The paper is organized as follows. In section \ref{sec:impl}, we
review the YFS algorithm, motivating and investigating the procedure
to include higher order corrections at a given perturbative order. 
In section \ref{sec:results}, we summarize the
numerical results for the decays $Z\to \ell^+\ell^-$, $W\to \ell\nu$ in Drell-Yan
production. There we also present results for $H\to \ell^+\ell^-$ decays in
hadronic Higgs production.  The measurement of the latter is highly
challenging due to small leptonic Higgs couplings but potentially
achievable at the HL-LHC.
We discuss and conclude in section \ref{sec:conclusions}.

\section{Implementation}
\label{sec:impl}

\subsection{The YFS formalism}
\label{sec:impl:yfs}

In this section, we will briefly recapitulate the YFS formalism in
a form appropriate for the approximate description of photon radiation
in particle decays, using the exponentiation of the universal soft
limit of matrix elements for real and/or virtual photons and its
systematic improvement through exact fixed-order calculations.
The decay rate of a decaying particle with mass $m$ and momentum $q$
into a set of decay products with momenta $p_f$, fully inclusive
with respect to the number of real and virtual photons
$n_R$ and $n_V$ reads
\begin{equation} 
\Gamma = \frac{1}{2m} \sum_{n_R = 0}^\infty
\frac{1}{n_R!} \int \mr{d}\Phi_p \mr{d}\Phi_k (2\pi)^4
\delta\left(q -\sum\limits_f p_f - \sum\limits_{i=0}^{n_R} k_i\right)
\left|\sum_{n_V=0}^\infty\mc{M}_{n_R}^{n_V+\frac{1}{2}n_R}\right|^2\,.
\end{equation}
Compared to the original, Born-level matrix element $\mc{M}_0^0$ describing
the decay, the matrix elements $\mc{M}_{i}^j$ include $i$ real photons at
the overall order $j$ in the electromagnetic coupling $\alpha$. This
equation for the decay rate describes an unrealistic situation, where we
are able to calculate all matrix elements, to all orders, and where we can
integrate them over their respective full phase space, while in reality 
at most the first few orders in perturbation theory can be calculated.
The YFS algorithm addresses this by dressing the lowest order matrix
elements with exponentiated eikonal factors that capture the leading
logarithmic behaviour of the amplitude, thus providing an all-order
description of QED radiation correct in this limit.  The full result is
restored, order by order in perturbation theory, by including the
subleading process-dependent parts of the amplitude.

Encapsulating the leading soft behaviour of a single virtual photon in a
process-independent factor $\alpha B$, the full one-loop matrix element
can be written as
\begin{equation} 
\mc{M}_0^1 = \alpha B \mc{M}_0^0 + M_0^1,
\end{equation}
where $M_0^1$ is an infrared subtracted matrix element including a
virtual photon.  Note that throughout this paper we assume all charged
particles to be massive; consequently the matrix elements do not
exhibit collinear singularities.  YFS showed that the simple structure
at first order extends also to all further virtual photon
corrections.
Including the appropriate symmetrisation prefactors this generalises to 
\begin{equation} 
\mc{M}_0^{n_V} = \sum_{n=0}^{n_V}M_0^{n_V-n}
\frac{\left(\alpha B\right)^n}{n!}.\label{sec:impl:eq:virtexp}
\end{equation}
Summing over all numbers of virtual photons $n_V$, we find that
the soft behaviour exponentiates,
\begin{equation} 
\sum_{n_V=0}^{\infty} \mc{M}_0^{n_V} =
\exp\left(\alpha B\right)\sum_{n_V=0}^{\infty} M_0^{n_V}.
\end{equation}
In QED, this argument generalises to matrix elements also containing any
number $n_R$ of real photons, {\it i.e.},
\begin{equation} 
\left|\sum_{n_V}^\infty
\mc{M}_{n_R}^{n_V+\frac{1}{2}n_R}\right|^2 = \exp\left(2\alpha
B\right)\left|\sum_{n_V=0}^{\infty}
M_{n_R}^{n_V+\frac{1}{2}n_R}\right|^2,
\end{equation}
where the $M_{n_R}^{n_V+\frac{1}{2}n_R}$ are free of virtual soft singularities,
but will still contain the soft divergences due to real photons.

In contrast to the virtual amplitudes, the factorization for real photons
occurs at the level of the squared matrix elements.  For a single photon
emission it reads
\begin{equation} 
\frac{1}{2\left(2\pi\right)^3}
\left|\sum_{n_V}^\infty M_{1}^{n_V+\frac{1}{2}}\right|^2 =
\tilde{S}\left(k\right) \left|\sum_{n_V}^\infty M_{0}^{n_V}\right|^2 +
\sum_{n_V=0}^\infty \tilde{\beta}_1^{n_V+1}\left(k\right)\,,
\end{equation}
where the eikonal factor $\tilde{S}\left(k\right)$ contains the real soft
divergence. We denote the complete infrared finite squared real matrix element
as $\tilde{\beta}_{n_R}^{n_V+n_R}$ and employ the abbreviation
\begin{equation} 
\tilde{\beta}_{n_R} = \sum_{n_V=0}^\infty
\tilde{\beta}_{n_R}^{n_V+n_R}
\end{equation}
to write the squared matrix element for the emission of $n_R$ real
photons, summed over all numbers $n_V$ of virtual photons as
\begin{align}
\begin{split} 
\left(\frac{1}{2\left(2\pi\right)^3}\right)^{n_R}
\left|\sum_{n_V}^\infty M_{n_R}^{n_V+\frac{1}{2}n_R}\right|^2 =&
\tilde{\beta}_0
\prod_{i=1}^{n_R}\left[\tilde{S}\left(k_i\right)\right] +
\sum_{i=1}^{n_R}\left[\frac{\tilde{\beta}_1\left(k_i\right)}{\tilde{S}\left(k_i\right)}\right]
\prod_{j=1}^{n_R}\left[\tilde{S}\left(k_j\right)\right] \\
& + \sum_{\substack{i,j=1 \\i\neq
j}}^{n_R}\left[\frac{\tilde{\beta}_2\left(k_i,k_j\right)}{\tilde{S}\left(k_i\right)\tilde{S}\left(k_j\right)}\right]
\prod_{l=1}^{n_R}\left[\tilde{S}\left(k_l\right)\right] + \dotsm +
\tilde{\beta}_{n_R}\left(k_1,\dotsm,k_{n_R}\right).
\end{split}
\end{align}
This expression contains all possible divergences due to real photon
emission in the eikonal factors. The first term describes the leading
logarithmic behaviour, and contains all virtual insertions to the
matrix element without any real photon emission through $\tilde{\beta}_0$.
The second term corrects the approximate expression in the $\tilde{S}$
for the real emission of one additional photon to the exact result,
and so on.  Expanding the $\tilde{\beta}_i$ in the electromagnetic
coupling constant $\alpha$ we can get a systematic, perturbative
expansion.  Demanding agreement with the exact results up to
$\mc{O}\left(\alpha^2\right)$ results in
\begin{align}
\begin{split} 
\left(\frac{1}{2\left(2\pi\right)^3}\right)^{n_R}
\left|\sum_{n_V}^\infty M_{n_R}^{n_V+\frac{1}{2}n_R}\right|^2 
=&
\left(\tilde{\beta}_0^0+
  \tilde{\beta}_0^1+\tilde{\beta}_0^2\right)\prod_{i=1}^{n_R}\left[\tilde{S}\left(k_i\right)\right]
+
\sum_{i=1}^{n_R}\left[\frac{\tilde{\beta}_1^1+\tilde{\beta}_1^2\left(k_i\right)}{\tilde{S}\left(k_i\right)}\right]
\prod_{j=1}^{n_R}\left[\tilde{S}\left(k_j\right)\right]\\
& +
\sum_{\substack{i,j=1 \\i\neq
    j}}^{n_R}\left[\frac{\tilde{\beta}_2^2\left(k_i,k_j\right)}{\tilde{S}\left(k_i\right)\tilde{S}\left(k_j\right)}\right]
\prod_{l=1}^{n_R}\left[\tilde{S}\left(k_l\right)\right] +
\mc{O}\left(\alpha^3\right)\,,
\end{split}
\end{align}
effectively making explicit the terms related to virtual photon
corrections\footnote{ For an agreement correct up to order
  $\mc{O}\left(\alpha\right)$, we would need to remove
  $\tilde{\beta}_0^2$, $\tilde{\beta}_1^2$ and $\tilde{\beta}_2^2$.
  By far and large this has already been implemented
  in~\cite{Schonherr:2008av}.}.

Completing the exponentiation of the leading logarithmic behaviour in
both real and virtual photon corrections and correcting the result to
the exact first order expression we arrive at
\begin{align}
\begin{split} 
  2m\cdot \Gamma =
  \int \mr{d}y \int
  \mr{d}\Phi_{p_f} \biggl\{&\exp\left(2\alpha B\right)\int
  \mr{d}y \exp\left[iy\left(q -\sum p_f\right)+\int
    \frac{\mr{d}^3k}{k} \tilde{S}(k) e^{-iyk}\right] \\
  &\times\left(\tilde{\beta}_0^0+\tilde{\beta}_0^1+
  \int\frac{\mr{d}^3k}{k} e^{-iyk}
  \tilde{\beta}_1^1(k)+ \mc{O}\left(\alpha^2\right)\right)\biggr\}.
\end{split}
\end{align}

In this expression, all virtual infrared singularities are contained in $B$
while all real infrared singularities are contained in the integral over
$\tilde{S}(k)$.  There, terms diverging in the limit $k\to 0$ can easily
be isolated by defining a small soft region $\Omega$ that contains the
limit $k\to 0$ such that $\Theta(k,\Omega) = 1$ if $k \notin \Omega$:
\begin{align}
\begin{split} 
  \int \frac{\mr{d}^3k}{k} \tilde{S}(k) e^{-iyk}  = &
  \int\frac{\mr{d}^3k}{k}\left\{\tilde{S}(k)
  \left[\left(\vphantom{\frac12}1-\Theta(k,\Omega)\right) +
    e^{-iyk} \Theta(k,\Omega) +
    \left(\vphantom{\frac12}e^{-iyk}-1\right)
    \left(\vphantom{\frac12}1-\Theta(k,\Omega)\right)\right]\right\} \\ 
  = & \vphantom{\frac12}
  2\alpha \tilde{B}(\Omega) + D(\Omega).
\end{split}
\end{align}
The two functions $\tilde{B}(\Omega)$ and $D(\Omega)$ are given by
\begin{align}
\begin{split} 
  2\alpha \tilde{B}(\Omega) =&
  \int \frac{\mr{d}^3k}{k}\tilde{S}(k) e^{-iyk}
  \left(\vphantom{\frac12}1-\Theta(k,\Omega)\right)\,,\\
  D(\Omega) =& \int \frac{\mr{d}^3k}{k} \tilde{S}(k)
  \left[e^{-iyk}\Theta(k,\Omega)+
    \left(\vphantom{\frac12}e^{-iyk}-1\right)
    \left(\vphantom{\frac12}1-\Theta(k,\Omega)\right)\right]\,,
\end{split}
\end{align}
where the former contains the infrared singularities and the latter is
infrared regular.  This separation allows the re-expansion of the
exponentiated integral and the re-instating of explicit momentum
conservation through $\delta$-functions, arriving at the master formula
for the decay rate in the YFS approach:
\begin{align}
\begin{split} 
  2m\cdot \Gamma = \sum_{n_R} \frac{1}{n_R!} \int
  \mr{d}\Phi_{p_f} \mr{d}\Phi_k & \left(2\pi\right)^4
  \delta^4\left(q-\sum\limits_f p_f-\sum\limits_{i=0}^{n_R} k_i\right)\\
  &\times e^{Y(\Omega,\,\{q\})}\prod_{i=1}^{n_R}
  \tilde{S}\left(k_i,\{q\}\right)\Theta\left(k_i,\Omega\right)
  \;\tilde{\beta}_0^0(\{q\})\;\mc{C}(\{p\},\{q\})
  \;\mc{J}(\{p\},\{q\}).\label{sec:impl:eq:master}
\end{split}
\end{align}
In the equation above we made the dependence on momenta explicit: the
Born-level momenta of the process before QED radiation are denoted by
$q_i$, while the momenta of the full final state including radiation
are labelled $p_i$.  The mapping between both sets of momenta is
detailed below.  The individual terms are
\begin{itemize}
\item the YFS form factor
  \begin{equation} 
    Y(\Omega) = \sum_{i<j} Y_{ij}(\Omega) =
    2\alpha\left(B_{ij}+\tilde{B}_{ij}(\Omega)\right),
  \end{equation}
  with the sum running over all pairs of charged particles and the
  soft factors given by
  \begin{align} 
    B_{ij} &= -\frac{i}{8\pi^3}Z_iZ_j\theta_i\theta_j\int
    \mr{d}^4k \frac{1}{k^2}\left(\frac{2q_i\theta_i-k}{k^2-2\left(k\cdot
      q_i\right)\theta_i} + \frac{2q_j\theta_j+k}{k^2+2\left(k\cdot
      q_j\right)\theta_j}\right)^2, \label{sec:impl:eq:B} \\
    \tilde{B}_{ij}\left(\Omega\right) &=
    \frac{1}{4\pi^2}Z_iZ_j\theta_i\theta_j\int \mr{d}^4k\ 
    \delta\left(k^2\right)
    \left(\vphantom{\frac12}1-\Theta\left(k,\Omega\right)\right)
    \left(\frac{q_i}{q_i\cdot
      k} - \frac{q_j}{q_j\cdot k}\right)^2.\label{sec:impl:eq:Btilde}
  \end{align}
  These two terms contain all infrared virtual and real divergences which
  cancel due to the
  Kinoshita-Lee-Nauenberg theorem~\cite{Kinoshita:1962ur,Lee:1964is},
  guaranteeing the finiteness of $Y(\Omega)$ and of the decay width. $Z_i$
  and $Z_j$ are the charges of the particles $i$ and $j$, and the factors
  $\theta = \pm 1$ for particles in the final or initial state,
  respectively.  We provide expressions for $B_{ij}$ in final-final and
  initial-final dipoles in terms of scalar master integrals in appendix
  \ref{sec:app:Infrared_Factors}. The calculation of the full form
  factor can be found in \cite{Schonherr:2008av};
\item the eikonal factor $\tilde{S}\left(k\right)$
  \begin{equation} 
    \tilde{S}\left(k\right) = \sum_{i<j}
    \tilde{S}_{ij}\left(k\right) =
    \frac{\alpha}{4\pi^2}\sum_{i<j}Z_iZ_j\theta_i\theta_j
    \left(\frac{q_i}{q_i\cdot k} - \frac{q_j}{q_j\cdot k}\right)^2
    \label{sec:impl:eq:S}
  \end{equation}
  describing the soft emission of a photon off a collection of charged
  particles;
\item the lowest order matrix element $\tilde{\beta_0^0}$;
\item a correction factor $\mc{C}$ to the full matrix element, given by
  \begin{equation} 
    \mc{C} = 1 +
    \frac{1}{\tilde{\beta}_0^0}
    \left(\tilde{\beta}_0^1 +
    \sum_{i=1}^{n_\gamma}
    \frac{\tilde{\beta_1^1}\left(k_i\right)}
         {\tilde{S}\left(k_i\right)}\right) +
    \frac{1}{\tilde{\beta}_0^0}\left(\tilde{\beta}_0^2 +
    \sum_{i=1}^{n_\gamma}
    \frac{\tilde{\beta_1^2}\left(k_i\right)}{\tilde{S}\left(k_i\right)}
    + \sum_{\substack{i,j=1 \\i\neq j}}^{n_\gamma}
    \frac{\tilde{\beta}_2^2\left(k_i,k_j\right)}
         {\tilde{S}\left(k_i\right)\tilde{S}\left(k_j\right)}\right)
         + \frac{1}{\tilde{\beta}_0^0}
         \mc{O}\left(\alpha^3\right). \label{sec:impl:eq:C}
  \end{equation}
  The terms in the first bracket describe the next-to-leading order
  (NLO), {\it i.e.}\ the $\mc{O}\left(\alpha\right)$ term of the
  expansion, and the terms in the second bracket describe the
  next-to--\-next-to-leading order (NNLO), {\it i.e.} the
  $\mc{O}\left(\alpha^2\right)$ term of the expansion.  In this
  publication, we will primarily be concerned with this correction
  factor, in particular with the virtual corrections at NLO, the term
  $\tilde{\beta}_0^1$, which we extend to an expression at full NLO in the
   electroweak theory for the decays of the weak bosons, as well
  as the complete NNLO bracket which we will be calculating for the
  neutral weak bosons;
\item and the Jacobean $\mc{J}$ capturing the effect of the momenta
  mapping.
\end{itemize}

The mappings relevant for particle decays
of both uncharged and charged initial particles have been outlined in
section 3.3 of \cite{Schonherr:2008av}, but we repeat them in Appendix \ref{sec:impl:MomMapping}
for the benefit of the interested reader.  

\subsection{Motivation for higher order corrections}
\label{sec:impl:motiv}

The previous subsection introduced the YFS procedure for dressing the lowest
order matrix element with soft radiation to all orders. This basic
procedure, in which $\mc{C} = 1$, yields photon distributions that are
correct in the limit of soft radiation. For the remainder of this
paper, we will call this the \textit{soft approximation}. Away from
the strict soft limit, exact matrix elements are necessary to describe
observables at the required accuracy, and we described the procedure
for their systematic incorporation. Hard photon radiation occurs
predominantly collinear to the emitter and more frequently in
processes with large energy-to-mass ratios of the involved
particles. With this in mind, generic collinear corrections for the
real matrix element, based on the splitting functions developed in
\cite{Dittmaier:1999mb}, were employed in \cite{Schonherr:2008av} to
account for hard QED radiation in the \textit{soft-collinear
  approximation}. While this approximation correctly describes
radiation in the limits of soft and collinear radiation, it does neither
account for  interference effects nor hard wide-angle
radiation. In order to capture these effects correctly, full matrix
elements for real and virtual photon radiation have to be added, some of
which have already been included in~\cite{Schonherr:2008av}.

\begin{figure}[t] \centering
\includegraphics[scale=\figscale]{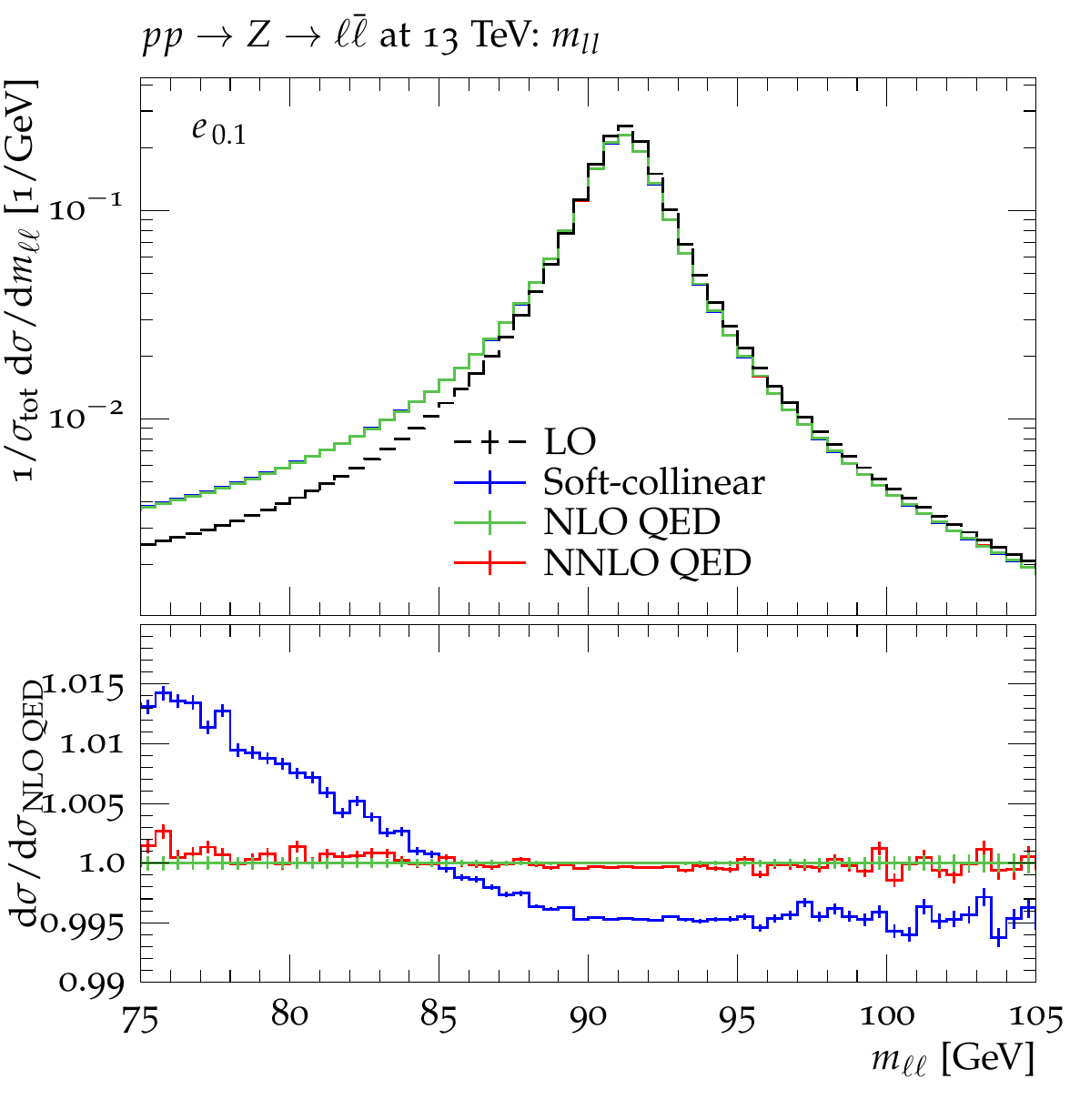}
\includegraphics[scale=\figscale]{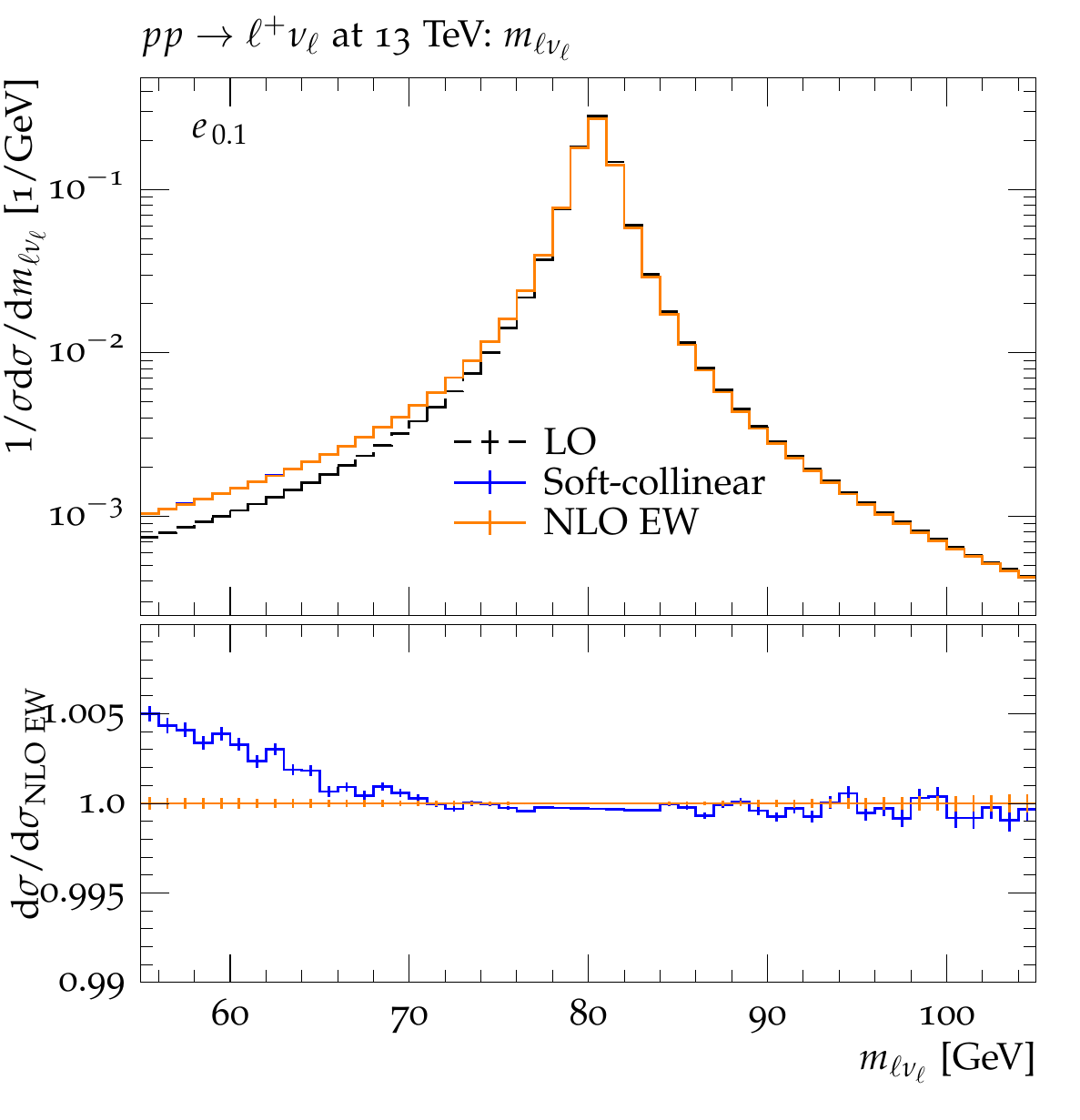}
\caption{The invariant mass $m_{\ell\ell}$ of the two leptons in $Z$-boson
  decays on the left and the invariant mass $m_{\ell\nu}$ of the charged
  lepton and the neutrino in $W$-boson decays on the right are shown
  for the processes $pp \to Z\to e^+ e^-$ and $pp \to W^+\to e^+
  \nu_e$ respectively. Different levels of fixed order accuracy are
  compared. The electrons in both cases are dressed with collinear
  photons within $\Delta R=0.1$.}
\label{fig:impl:motiv}
\end{figure}

For illustration, in Fig.\ \ref{fig:impl:motiv} (left) we compare the soft-collinear, the 
NLO QED-correct and the NNLO QED-correct results for the invariant
mass $m_{\ell\ell}$ of the electrons produced in $Z$-boson decays.
To guide the eye we also show the leading-order result.
The NLO QED result represents the maximal accuracy of the
implementation in \Photons as described in \cite{Schonherr:2008av}.
Fig.\ \ref{fig:impl:motiv} (left) clearly shows the necessity to include photon
radiation in the first place. Photon radiation causes a significant
shape difference, shifting events from large to lower $m_{\ell\ell}$. 
This effect is a lot more striking in the decay into the
lighter leptons, such as for the electrons exhibited here, which are much
more likely to radiate photons. We can also appreciate that while the
soft-collinear approximation does a good job of describing the
distribution near the peak, it predicts a harder spectrum at lower
values of $m_{\ell\ell}$.
The peak region corresponds to the
limit of soft photon radiation, while the latter region corresponds to
hard photon radiation. This observation thus suggests that in order to
capture the behaviour of the distribution over its entirety, we need
to employ full matrix elements. It is then natural to ask whether
higher order corrections beyond the NLO in QED are required as
well. The impact of NNLO QED corrections is already illustrated in 
Fig.\ \ref{fig:impl:motiv} (left) and the description of these 
and of full NLO EW corrections will be the focus of the next two subsections.

\subsection{NLO electroweak corrections}
\label{sec:impl:ew}

The discussion in Section \ref{sec:impl:yfs} was restricted to QED corrections only.
Since the exponentiation relies on the universal behaviour
of the amplitudes in the soft limit only, additional fixed-order corrections
can easily be added, as long as they are \textit{not} divergent in the
soft limit and thus do not spoil the soft-photon exponentiation.  This is,
in fact, the case for the weak part of the corrections in the full
electroweak theory, where the masses of the weak bosons regulate the soft
divergence that is plaguing the massless photon. In this work, we are
concerned with the decays of weak bosons; consequently, there is no phase
space available for the emission of a real, massive weak boson, and the
additional electroweak corrections contribute only to the virtual corrections
$\tilde{\beta}_0^{n_V}$.

The known one-loop virtual corrections for the decays of the
electroweak bosons~\cite{Hollik:1988ii,Denner:1991kt} have been
implemented in a number of programs dedicated to electroweak precision
calculations already mentioned in the introduction.  They can be
calculated analytically with programs such as
\FeynCalc~\cite{Mertig:1990an,Shtabovenko:2016sxi},
\FormCalc~\cite{Hahn:1998yk} or \PackageX~\cite{Patel:2015tea}, and
numerically with programs such as
\GoSam~\cite{Chiesa:2015mya,Chiesa:2017gqx}, 
\MadGraphFive~\cite{Alwall:2014hca,Frederix:2018nkq},
\OpenLoops~\cite{Cascioli:2011va,Kallweit:2014xda} or
\Recola~\cite{Actis:2012qn,Actis:2016mpe}.  The two-loop virtual
electroweak corrections are not fully known yet, with only partial
results for particular observables available, see for
example~\cite{Dubovyk:2016aqv,Dubovyk:2018rlg}.

We implemented the electroweak corrections for the decays $Z\to \ell^+\ell^-$,
$H\to\ell^+\ell^-$ and $W\to\ell\nu$ in the YFS correction factor $\mathcal{C}$. 
In doing so, we also re-implemented and
re-validated the QED corrections in a more straightforward way. 
In our calculation we retain
the full dependence on the lepton masses in the decay $H\to\ell^+\ell^-$.
In the two other decays we keep them only in the QED corrections, where
they are required to regularize the collinear singularities, while we
neglect them in the other contributions. To this end we used the
vertex form factors found in~\cite{Bardin:1999ak} to describe the
virtual corrections to the vertices. We renormalize the theory using
the on-shell renormalization scheme, following the treatment described
in~\cite{Denner:1991kt}.
We have validated the amplitudes on a point-by-point level against an
implementation in \OpenLoops \cite{Cascioli:2011va,Kallweit:2014xda},
all in the case of massless leptons for $Z$- and $W$-boson decays, and
for the case of a Higgs decay into massive fermions. In addition, we
also validated the values of the scalar integrals including masses
against
\Collier~\cite{Denner:2002ii,Denner:2005nn,Denner:2010tr,Denner:2016kdg}
and \QCDLoop~\cite{Carrazza:2016gav}, as well as individual renormalization
constants for massive leptons against \OpenLoops. Real corrections due
to the emission of an additional photon are calculated in the helicity
formalism~\cite{Kleiss:1985yh,Ballestrero:1992dv,Ballestrero:1994jn}
using building blocks available within \Sherpa~\cite{Krauss:2001iv}.
We validated these corrections explicitly, against
\WZGRAD~\cite{Baur:1998kt,Baur:2001ze,Baur:2004ig} and by internal
comparison with \Amegic~\cite{Krauss:2001iv} and
\Comix~\cite{Gleisberg:2008fv}.  We also validated full cross sections
against \WZGRAD~\cite{Baur:1998kt,Baur:2001ze,Baur:2004ig}.

For the decays of $Z$- and Higgs-bosons, we further implement an
option including only QED corrections. In the decay of neutral bosons,
this choice forms a gauge--\-invariant subset of the full electroweak
corrections and can thus be considered independently. Practically,
this amounts to turning off the purely weak vertex form factors as
well as turning off those parts of the renormalization constants that
are of weak origin. This option is not available in the case of a
$W$-boson decay as the $W$ itself couples to the photon. We list the
relevant form factors, renormalization constants and the necessary
modifications in the pure QED case in Appendix~\ref{sec:app:NLO_EW}.

As an illustration in Fig.\ \ref{fig:impl:motiv} (right) we compare the LO, the soft-collinear 
and the full NLO-correct results for the invariant of the charged electron and the neutrino in $W$-boson
decays. As for the Z decay, the inclusion of the exact fixed-order corrections is mandatory for 
a reliable prescription below the resonance peak.

\subsection{NNLO QED corrections}
\label{sec:impl:nnlo}

We will now turn to the discussion of  NNLO QED corrections to $Z$-
and Higgs-boson decays. 
 They comprise double-virtual, real-virtual and real-real
contributions. The NNLO QED corrections can be combined with the full
NLO EW corrections, and we will label this combination as ``NNLO QED
$\oplus$ NLO EW''. As illustrated in Fig.\ \ref{fig:impl:motiv} (left) the NNLO corrections yield
very small corrections beyond NLO - at least in observables defined at LO.
However, their inclusion ensures precision at the sub-percent level required for
future Drell-Yan measurements.

\subsubsection{Double virtual corrections}
\label{sec:impl:VV}

The two-loop pure QED corrections to the form factor for the $Z$-boson
decay have been known in the limit of small lepton masses since the
LEP era~\cite{Burgers:1985qg,Berends:1987ab}. 
Including full mass dependence, currently  the
two-loop QED form factor is only known for the decay of a virtual photon~\cite{Bonciani:2003ai}.

To the best of our
knowledge, no QED two-loop form-factors are available for the decay of Higgs bosons. 
In principle they could be obtained from corresponding QCD results~\cite{Bernreuther:2004ih,Bernreuther:2004th,Bernreuther:2005rw,Bernreuther:2005gw,Ablinger:2017hst}
However, for simplicity here we
rely on the leading logarithmic behaviour only, $\tilde{\beta}_0^2 =
\frac{1}{2} \log^2\left(\frac{s}{m^2}\right)$. We find that for the
decays into bare muons, this is a sufficient approximation.  Appreciable effects 
due to this approximation might only be noticeable in Higgs decays into $\tau$-leptons.

For the decay of  $Z$-bosons, we use the results in Eqs. (2.15)
and (2.22) of~\cite{Berends:1987ab}, together with the subtraction
term $B$ expanded in the limit $s \gg m^2$. The results for
the form factors given in~\cite{Berends:1987ab} are sufficient as we
only require the squared contribution $\mr{Re}({M_0^2M_0^{0 \ast}})$. 
In fact, here the two-loop amplitude $M_0^2$ factorizes into a simple
factor multiplying the leading order matrix element.

The double virtual corrections can be decomposed, following the
procedure described in section \ref{sec:impl:yfs}, as
\begin{equation} \mathcal{M}_0^2 = M_0^2 +\alpha B M_0^1 +
\frac{(\alpha B)^2}{2!} M_0^0,
\end{equation}
such that the infrared subtracted matrix element reads
\begin{align}
  \begin{split}
    M_0^2 = \mathcal{M}_0^2 - \alpha B M_0^1 -
    \frac{(\alpha B)^2}{2!} M_0^0 
    = \mathcal{M}_0^2 - \alpha B \mathcal{M}_0^1 +
    \frac{(\alpha B)^2}{2!} M_0^0,
  \end{split}
\end{align}
using the decomposition $M_0^1 = \mathcal{M}_0^1
-\alpha B M_0^0$.
Employing the results of \cite{Berends:1987ab} and the form of the subtraction
term given in \EqRef{sec:app:Infrared_Factors:eq:photonmass}, we obtain
\begin{align}
  \begin{split} 
    \mr{Re} M_0^2 =  \frac{\alpha^2}{\pi^2}
    \biggl[\frac{1}{8}L^2+ L\left(-\frac{5}{32} -\frac{\pi^2}{8}
      +\frac{3}{2} \zeta (3)\right) 
      -\frac{9}{4}\zeta (3)
      -\frac{\pi^4}{15}+\frac{3}{2}+\pi^2\left(\frac{13}{32}-\frac{\log
        2}{2}\right)\biggr] M_0^0\,,
  \end{split}
\end{align}
where $\zeta(n)$ is the Riemann Zeta function, with
$\zeta(3) \approx 1.202056903159594$ and $L = \log(s/m^2)$.

The final correction term $\tilde{\beta}_0^2$ yields:
\begin{align}
  \begin{split}
    \tilde{\beta}_0^2 =&  M_0^2M_0^{0 \ast} + M_0^1M_0^{1
      \ast} +M_0^0M_0^{2 \ast} \\
    =&\frac{\alpha^2}{\pi^2}\biggl[\frac{1}{2}L^2
      + L\left(-\frac{13}{16} -\frac{\pi^2}{4}
      +3 \zeta (3)\right) 
      -\frac{9}{2}\zeta (3)
      -\frac{2\pi^4}{15}+\frac{13}{4}+\pi^2\left(\frac{17}{16}-\log
      2\right)\biggr] \tilde{\beta}_0^0\,.
  \end{split}
\end{align}

\subsubsection{Real-virtual corrections}
\label{sec:impl:RV}
The real-virtual corrections correspond the virtual corrections to the
process $X\to f\bar{f}^{(')}\gamma$, with one real, hard photon.  We can
write the infrared subtracted, squared real-virtual matrix elements as
\begin{equation} 
\tilde{\beta}_1^2\left(k_1\right) =
\frac{1}{2\left(2\pi\right)^3}\sum_{s_i,\lambda_j}\left(M_1^{\frac{3}{2}
\ast}\mc{M}_1^{\frac{1}{2}}+\mc{M}_1^{\frac{1}{2}
\ast}M_1^{\frac{3}{2}}\right) - \tilde{S}\left(k_1\right)
\tilde{\beta}_0^1,\label{eq:impl:RV}
\end{equation}
where $k_1$ denotes the momentum of the hard photon, and the sum in
the first term runs over the spins $s_i$ of the leptons and the
polarizations $\lambda_j$ of the vector bosons. The factor
$\tilde{S}\left(k_1\right)$ is calculated using the momenta mapped to
the single photon final state taking $k_1$ as the hard photon
momentum. For consistency, $\tilde{\beta}_0^1$ contains only the
one-loop QED corrections.
Using
\FeynCalc~\cite{Mertig:1990an,Shtabovenko:2016sxi} we rewrite the
amplitudes in terms of standard matrix elements multiplied by
expressions involving scalar master integrals. We have encoded the
neccessary master integrals
using~\cite{Denner:1991kt,Bardin:1999ak,Ellis:2007qk}.  We also use
the algorithm proposed in~\cite{Frellesvig:2016ske} for the evaluation
of the complex dilogarithm occuring in the master integrals.
We have confirmed the analytical
cancellation of the UV divergences upon including the renormalization
terms as well as the cancellation of the virtual IR divergences upon
inclusion of the infrared subtraction term.  However, the very nature
of the expressions involved increases the likelihood of numerical
instabilities in the evaluation of particular phase space points:
while strictly finite, separate terms in the expression may suffer from 
numerical instabilities, causing incomplete cancellations between
different terms.  The
reasons are twofold, and connected with the collinear regime of the
emissions:
\begin{itemize}
\item The YFS formalism relies on fermion masses to regularize the
  collinear singularities, which in the case of small fermion masses may
  amount to the evaluation of expressions very close to logarithmic singularities, 
  of the type $\log(s_{ij}/m^2)$, where $s_{ij} = (p_i+p_j)^2$
  is the invariant mass of two momenta. We find that in our implementation
  the amplitudes for the decays into electrons and to some extent also
  into muons are affected by numerical instabilities while the amplitudes
  for the decays into $\tau$'s are well-behaved.
\item In addition, the employed Passarino-Veltmann reduction may lead to the
  appearance of small Gram determinants in denominators.  One way to
  circumvent this issue is by employing an expansion in the Gram determinant
  for the problematic tensor integrals rather than the full reduction, as
  implemented for example in \Collier~\cite{Denner:2016kdg}. Since this
  requires the implementation of a significant number of expressions for
  different combinations of arguments in the tensor integrals and thus
  amounts to a large overhead, this is not pursued in this work.
\end{itemize}
To cure both problems, we instead use the following algorithm: We call
a phase space point ``collinear'' when $s_{ik} < a\cdot m_i^2$, where
$s_{ik}$ is the invariant mass between the photon and one of the
fermions in the process and $a$ is some predefined cutoff. Such a
phase space point will not be evaluated using the full matrix element
but rather using the quasi-collinear limit of the amplitude. Using
this limit, the calculational complexity of the amplitude is
significantly reduced and numerical instabilities are avoided.
As an additional rescue system, in case a bad phase space point should
still pass to be evaluated using the full matrix element, we also
check the scaling behaviour of the amplitude under a rescaling of all
dimensionful quantities. The expressions for the coefficients of the
master integrals can be rewritten using reduced quantities, {\it i.e.} all
dimensionful quantities are divided by the centre of mass energy of
the decay. In this way, dimensionful quantities only survive in the
master integrals themselves as well as in a single factor multiplying
the master integral.
The mass dimension of a four point function in four dimensions is 0, such
that upon rescaling by a common factor $\xi \neq 1$, the full expression
should remain unchanged, $\mc{M}(\xi) = \mc{M}(1)$. Different terms in the
matrix elements scale differently due to the different scaling behaviours of
the master integrals, so a deviation from the expected scaling
behaviour indicates numerical instabilities in the expression. For
 $\left|\mc{M}(1)/\mc{M}(\xi) - 1 \right| > c$, with $c$
some predefined cutoff, we set the real-virtual matrix element to 0.
We varified that all results remain unchanged varying the technical parameters a,~c.

\subsubsection{Real-real corrections}

The real-real corrections stem from the emission of two hard photons.
 For the implementation, we choose the same strategy as in the
case of single real corrections, by using helicity amplitudes and building
blocks already present in \Sherpa.
After setting up the amplitude like this, we can calculate the
infrared subtracted matrix element squared that enters into the
correction factor $\mc{C}$:
\begin{equation} 
\tilde{\beta}_2^2\left(k_1,k_2\right) =
\left(\frac{1}{2\left(2\pi\right)^3}\right)^2\sum_{s_i,\lambda_j}\mc{M}_2^{1
\ast}\mc{M}_2^1 - \tilde{S}\left(k_1\right)
\tilde{\beta}_1^1\left(k_2\right) - \tilde{S}\left(k_2\right)
\tilde{\beta}_1^1\left(k_1\right) - \tilde{S}\left(k_1\right)
\tilde{S}\left(k_2\right) \tilde{\beta}_0^0.
\end{equation}
In this formula, the $k_1$ and $k_2$ denote the momenta of the two
hard photons, the sum in the first term runs over the spins $s_i$ of
the leptons and the polarizations $\lambda_j$ of the vector
bosons. The $\tilde{S}\left(k_i\right)$ are calculated using the
momenta in the mapped $(n+2)$-dimensional phase space, using the pair
$k_1,k_2$ as the hard photons, see Appendix~\ref{sec:impl:MomMapping}.

\section{Results}
\label{sec:results}

\subsection{Setup}
\label{sec:results_setup}

In this section we present the numerical effects induced by the NLO EW
and NNLO QED corrections presented in the previous section, focussing
on the decays $Z\to\ell^+\ell^-$, $W\to\ell\nu$ and $H\to\ell^+\ell^-$ with
$\ell=\{e,\mu,\tau\}$ following hadronic neutral-current and
charged-current Drell-Yan and Higgs production respectively.

\begin{table} \centering
\begin{tabular}{ c | c | c }  & Mass [GeV] & Width [GeV] \\ 
\midrule
$Z$ & 91.1876 & 2.4952 \\ 
$W$ & 80.385 & 2.085 \\
$H$ & 125 & 0.00407 \\
$e$ & $0.511 \MeV$ & - \\
$\mu$ & $0.105 \GeV$ & - \\
$\tau$ & $1.777 \GeV$ & - \\
\hline
$1/\alpha\left(0\right)$ & 137.03599976 & 
\end{tabular}
\caption{Electroweak input parameters: gauge- and Higgs boson masses and widths, lepton masses and the EW coupling in the $\alpha(0)$ scheme. } 
\label{sec:results:tab:EW_parameters}
\end{table}

The results presented here are based on an implementation in the
\Photons module~\cite{Schonherr:2008av} of the \Sherpa Monte Carlo
framework (release version 2.2.4). We consider hadronic collisions at
the \LHC at 13\,TeV for the production of $Z$-, $W$- and Higgs-bosons
and their subsequent decays.  In the neutral-current Drell-Yan case we
require $65\GeV < m_{\ell\ell} < 115\GeV$, while for the other modes
no generation cuts are applied.  Since we aim to purely focus on the
effects of photon radiation in the decays, we turn off the QCD shower,
fragmentation and underlying event simulation. We use \Rivet 2.5.4
\cite{Buckley:2010ar} for the analysis.  For the case of electrons in
the final state, we perform the analysis either using bare leptons or
using dressed leptons with a radius parameter $\Delta R = 0.1$ or
$\Delta R = 0.2$. For the case of muon and $\tau$ final states only
bare results are shown.  We focus our results on a few key
distributions and always normalize to the respective inclusive cross
section.  Overall, we choose to focus on ratios between different
predictions, in order to highlight small subtle differences relevant
for precision Drell-Yan and Higgs physics.

Input parameters for the numerical results are chosen as listed in
Tab.\ \ref{sec:results:tab:EW_parameters}. The weak coupling $\alpha$
is defined in the on-shell $\alpha(0)$ scheme. This choice is sensible
as we are explicitly also investigating distributions in resolved
final-state photons. At the same time, the YFS formalism is strictly
only defined in the limit of soft photon emissions. In this input  scheme,
the sine of the weak mixing angle is a derived quantity $s_W^2 =1-
\frac{M_W^2}{M_Z^2}$. Gauge- and Higgs-boson  
widths are taken into account in a fixed-width scheme.

In the decays of $W$ and $Z$ bosons, we apply an IR technical cutoff
in the YFS formalism of $E_{\gamma, \text{cut}} = 0.1 \GeV$, while in
the Higgs-decay we reduce this value to $E_{\gamma, \text{cut}} = 0.01
\GeV$ in order to improve the resolution near the
resonance\footnote{It should of course be noted that the SM Higgs has
  a resonance width of only $\sim 4 \MeV$, which is smaller than this
  photon cut, suggesting that we still do not resolve the resonance
  well with this cut. However, we find that a cut of the order $10
  \MeV$ is necessary in order to guarantee a good performance of the
  method in both decay channels.  In any case, this smaller choice of
  the cutoff still allows a closer investigation of the regions close
  to the resonance in plots generated from the lepton momenta, as long
  as the binning is not chosen too fine. In particular the regions
  that will be populated through the radiation of photons from leptons
  in the resonance region will be included in this description.}. In
both cases, we keep an analysis cut of $E_{\gamma} > 0.1 \GeV$ for
observables involving photons.

\subsection{Neutral-current Drell-Yan  production}
\label{sec:results:Z}

\begin{figure}[t] \centering
\includegraphics[scale=0.55]{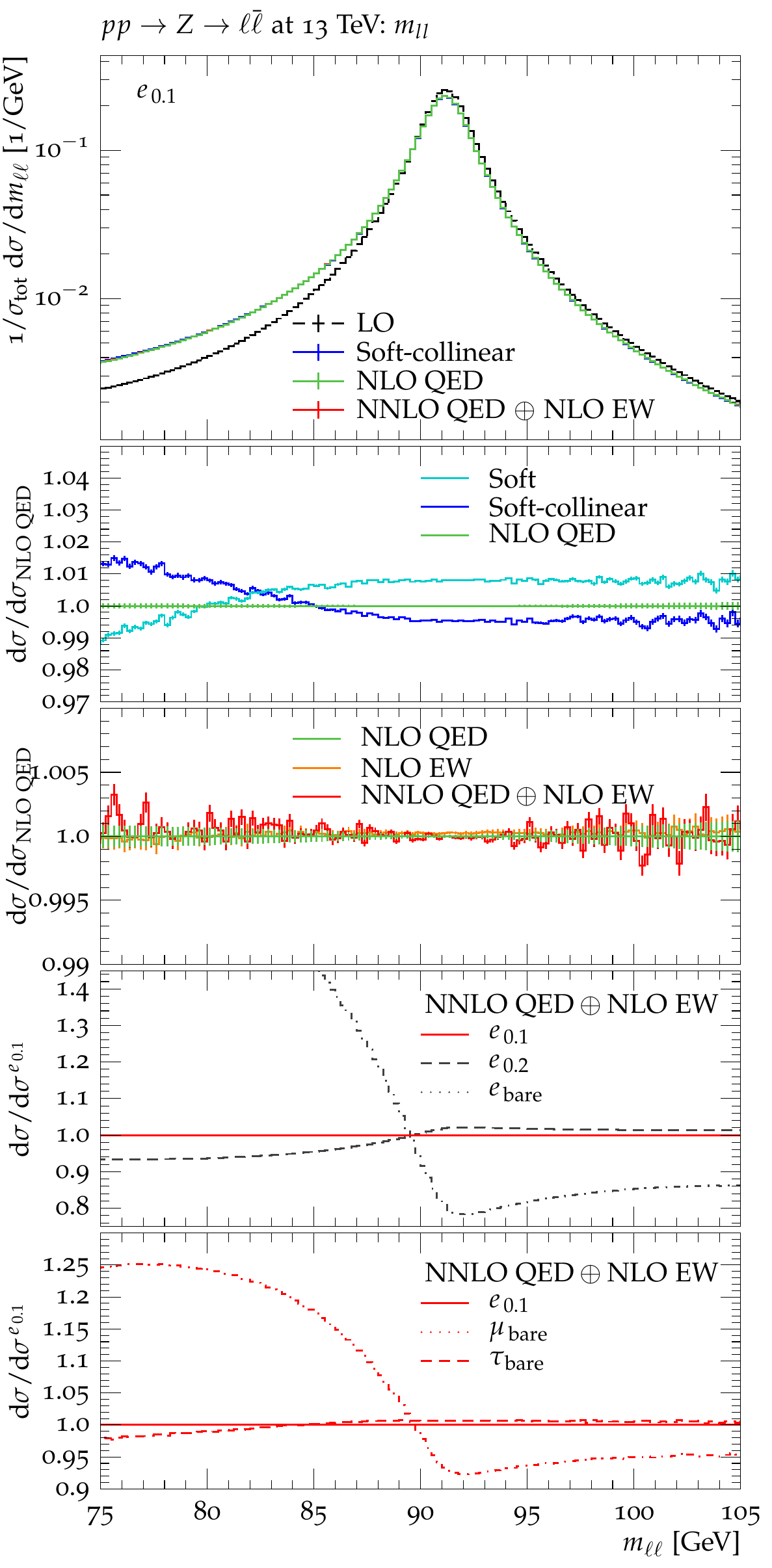}
\includegraphics[scale=0.55]{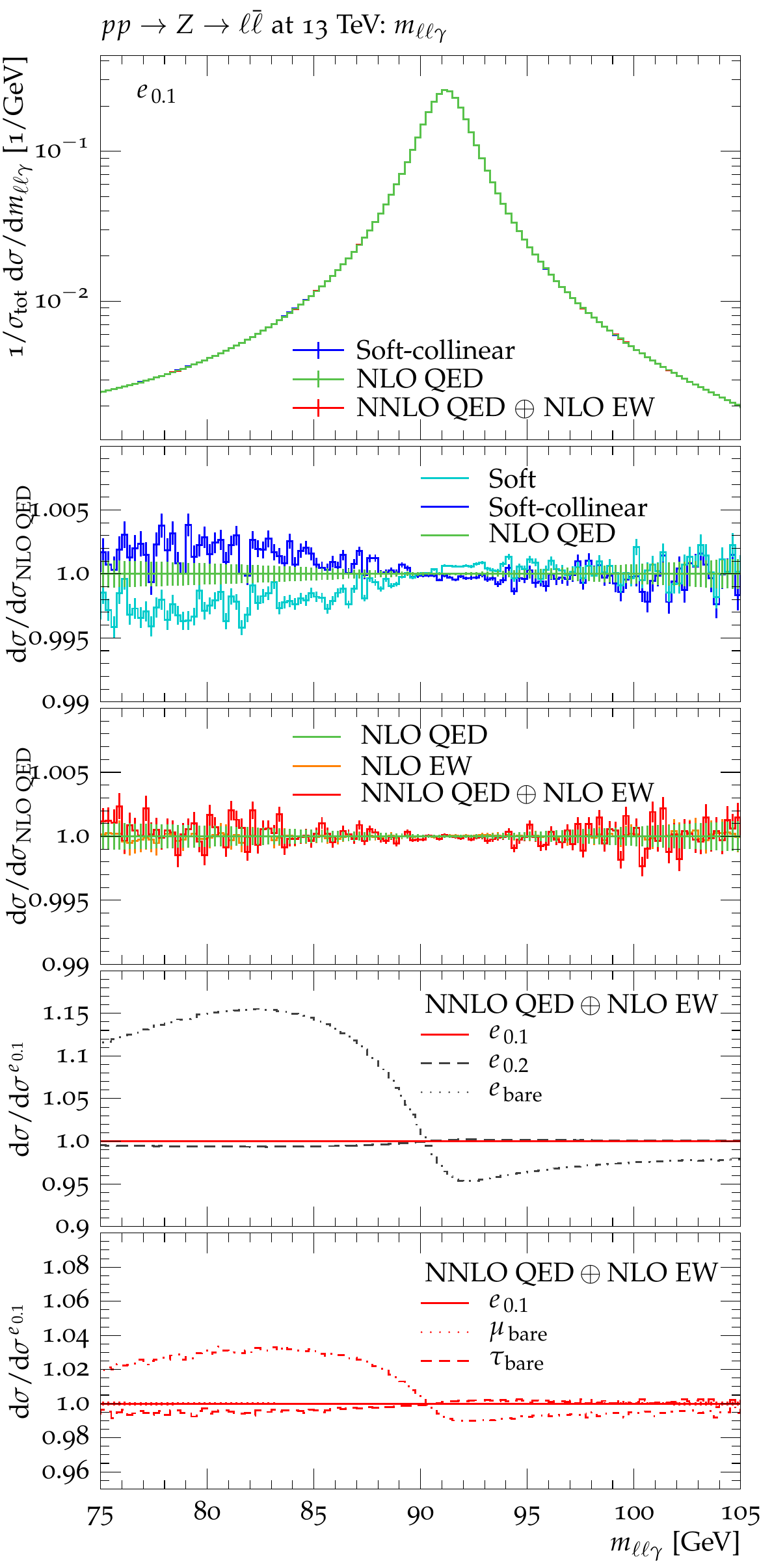}
\caption{On the left the invariant mass of the two leptons, $m_{\ell\ell}$,
  and on the right the invariant mass of the system of the two decay
  leptons and the closest photon, $m_{\ell\ell\gamma}$, is shown for $pp \to
  Z\to \ell^+ \ell^-$ production. Nominal predictions are shown for $pp \to
  Z\to e^+ e^-$ at LO, in soft-collinear NLO approximation, at NLO QED
  and at NNLO QED $\oplus$ NLO EW, where electrons are always dressed
  with collinear photons within $\Delta R=0.1$. The ratio plots highlight
  the effect of the considered higher-order corrections and the effect
  due to different photon dressing or lepton identity. See text for
  details.  }
\label{fig:results:Z:mll_mllg}
\end{figure}

\begin{figure}[t] \centering
\includegraphics[scale=\figscale]{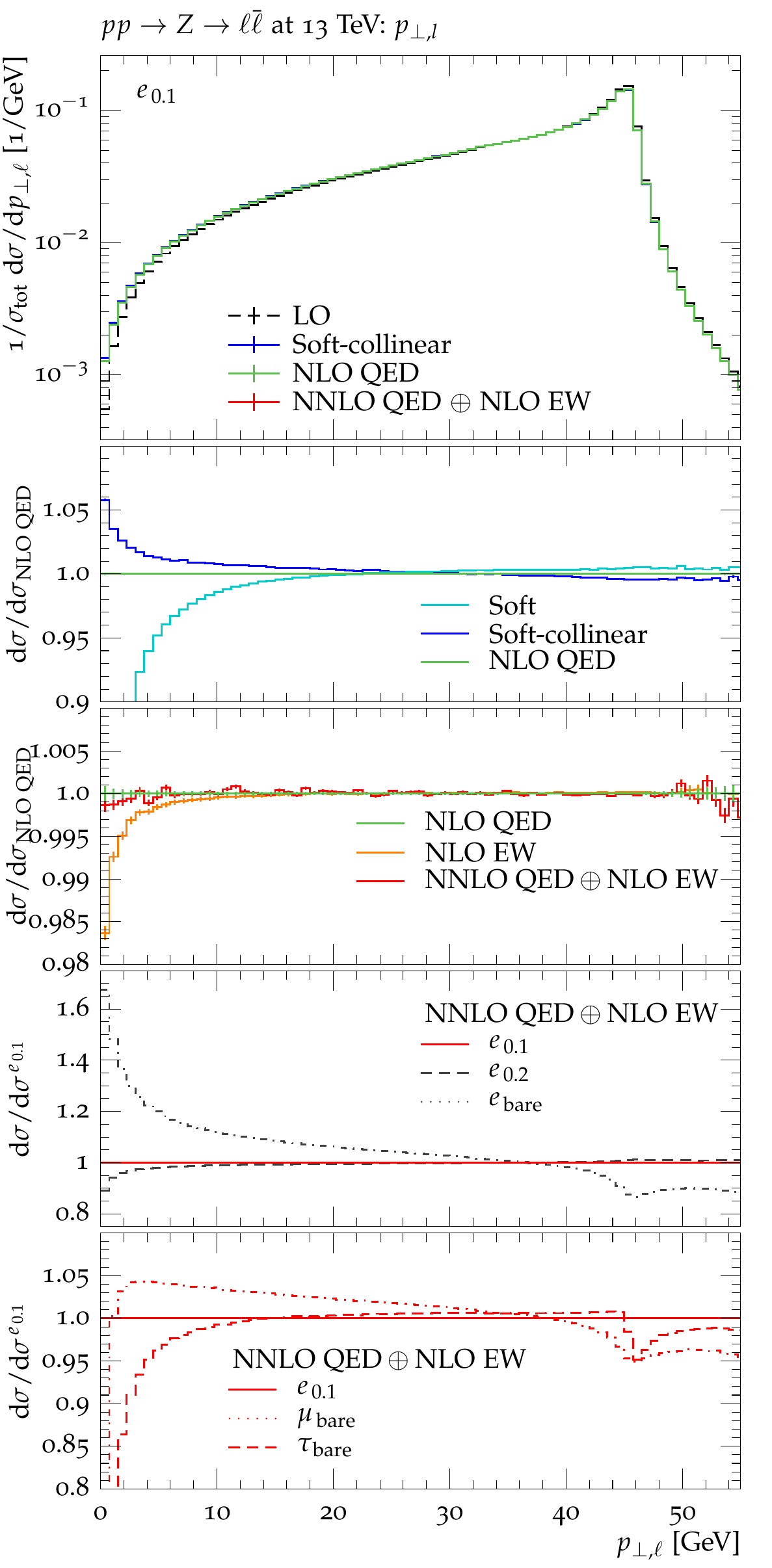}
\includegraphics[scale=\figscale]{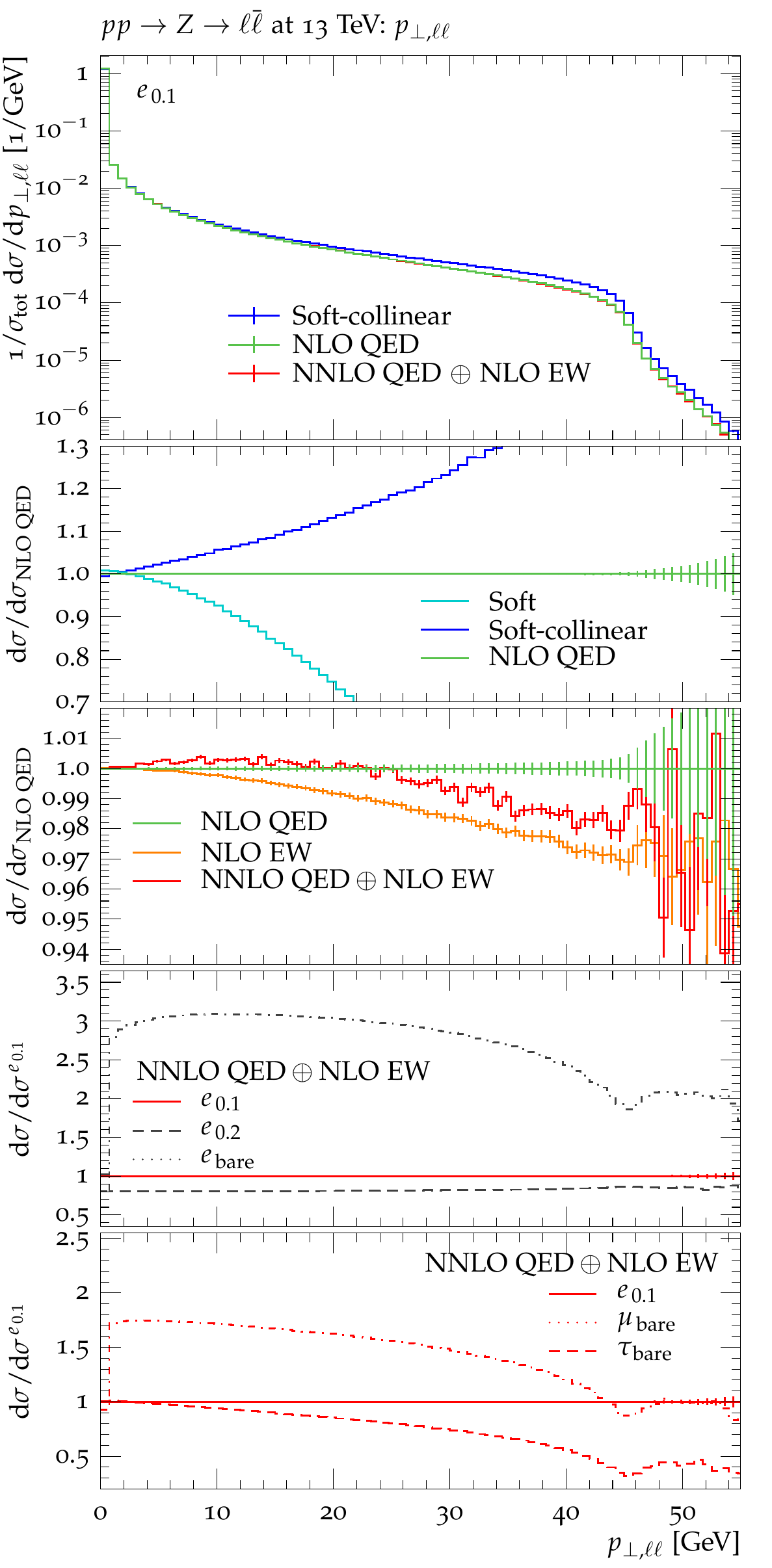}
\caption{Plots of the transverse momentum of the leptons,
  $p_{\perp,\ell}$, on the left and the transverse momentum of the system
  of the two decay leptons, $p_{\perp,\ell}$, on the right. Predictions
  and labels as in Fig.~\ref{fig:results:Z:mll_mllg}. }
\label{fig:results:Z:pTl_pTll}
\end{figure}

\begin{figure}[t] \centering
\includegraphics[scale=\figscale]{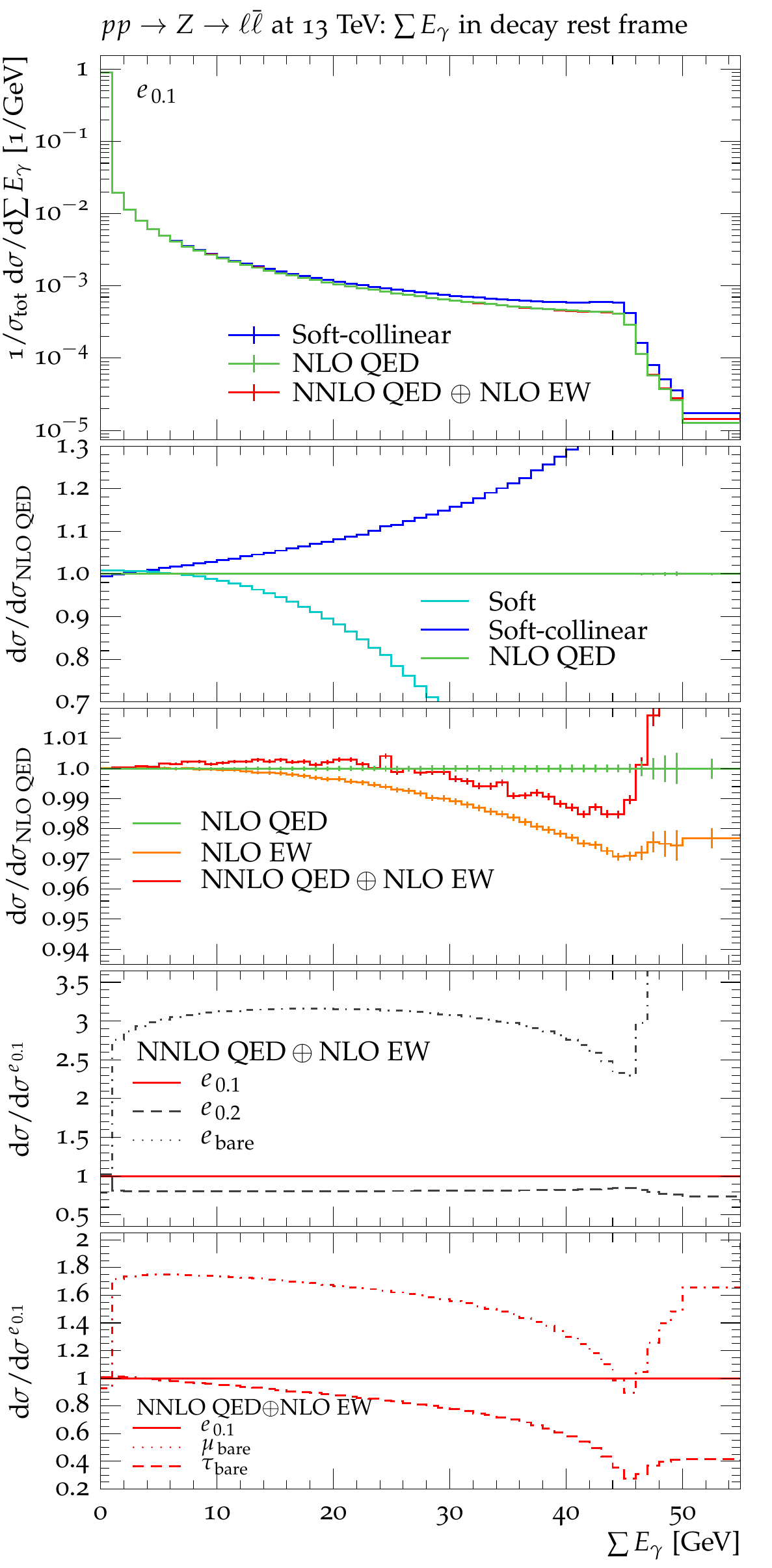}
\includegraphics[scale=\figscale]{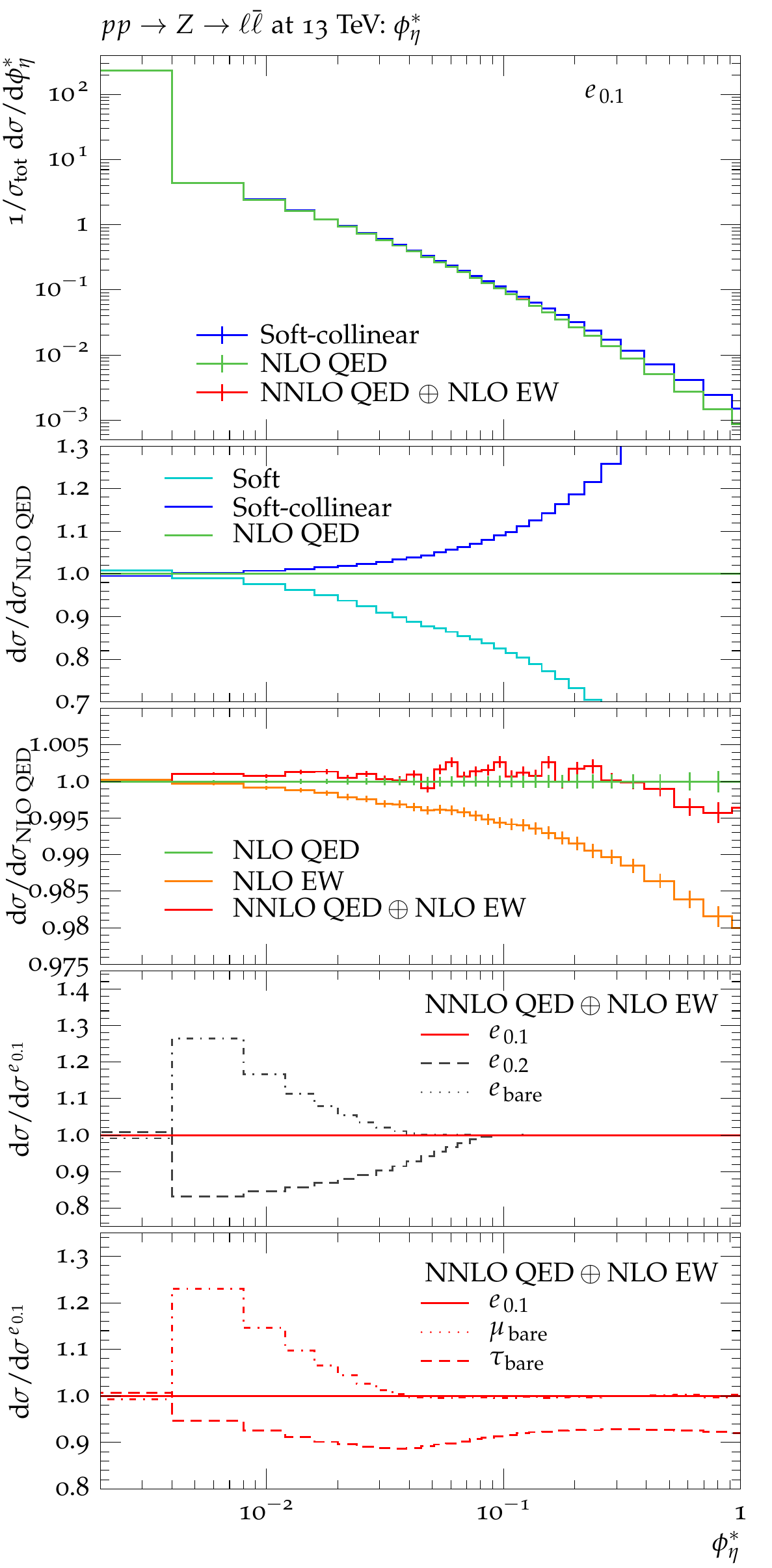}
\caption{Plots of the sum of the photon energies in the decay rest
  frame, $\sum_{n_{\gamma}} E_{\gamma}$, on the left and the
  $\phi_{\eta}^{\ast}$ variable on the right. Predictions and labels
  as in Fig.~\ref{fig:results:Z:mll_mllg}. }
\label{fig:results:Z:SumEg_Phistar}
\end{figure}

In
Figs.\ \ref{fig:results:Z:mll_mllg}-\ref{fig:results:Z:SumEg_Phistar}
we present several key observables in neutral-current Drell-Yan
production including higher-order QED corrections up to NNLO and EW
corrections up to NLO. All distributions are normalized and the
effects of the higher-order corrections typically manifest themselves
as very subtle shape distortions in the considered observables.  All
figures are identically structured and we show nominal predictions for
dressed di-electron production, i.e.\  collinear photon--electron pairs
with $\Delta R<0.1$ are combined, at LO (black), considering soft-collinear
QED corrections (blue), NLO QED corrections (green), and our best
predictions at NNLO QED $\oplus$ NLO EW (red).  In the first two
ratio plots we compare the predictions at NLO QED against the soft and
soft-collinear approximations and against the NLO EW and NNLO QED
$\oplus$ NLO EW predictions respectively.  In the third ratio plot we
investigate different dressing prescriptions of the electrons,
considering $\Delta R=0.2$ and undressed bare electrons. Finally, in the last
ratio plot we compare predictions for dressed electron with
corresponding ones for bare muons and $\tau$'s. In the latter two
ratios plots all predictions correspond to the most accurate level,
i.e.\ NNLO QED $\oplus$ NLO EW.

In Fig.\ \ref{fig:results:Z:mll_mllg}, we present the distributions of
the invariant mass of the two leptons (left) and of the invariant mass
of the system made up of the decay leptons and the photon closest to
either of them (right).
Already from the plots in Section \ref{sec:impl:motiv}, it is clear
that the inclusion of photon radiation is crucial for a reliable
description of the dilepton invariant mass. All higher-order
corrections significantly differ from the LO prediction, which fails
to describe the lineshape below the peak. At the NLO QED level
corrections beyond the soft and soft-collinear approximations induce
distortions up to the 1\% level.  In fact, the soft approximation does
not generate enough hard radiation, while the soft-collinear
approximation produces about 1\% too many events at low $m_{\ell\ell}$,
i.e.\ it seems to generate too much hard photon radiation. In this
observable both the NLO EW and NNLO QED corrections provide only a
marginal effect on the order of permille, and neither of these
corrections provides a significant shift of the peak of the
distribution.
Clearly, the dressing of the electrons has a significant effect on
this distribution, reflecting the sensitivity to QED radiation.  Bare
electrons show a significant shape difference compared to dressed
electrons. The results based on different dressing parameters however
differ by at most a few \%, suggesting that much of the photon
radiation occurs close to the electron.  Comparing different lepton
species, we see that muons, in comparison to the dressed electrons,
radiate significantly more, yielding up to 25\% more events at low
$m_{\ell\ell}$. In contrast, the heavier $\tau$'s radiate less in
comparison, resulting in differences with respect to dressed electrons
of only a few \%.

A very similar behaviour can be found in the invariant mass of the
dilepton system combined with the closest photon. As this observable
requires the emission of at least one photon, the NLO QED curve
corresponds effectively to a LO prediction. However, also the soft and
soft-collinear approximations describe this observable reasonably well
and higher order NNLO QED or NLO EW corrections are
negligible. Comparing the dressing parameters, we find much smaller
differences here: bare electrons only differing by about 15\% from the
dressed versions. There is barely a difference between the two
dressings. In the same manner, the difference between lepton species
is subdued as well: muons differing up to 2\% at most from dressed
electrons.

In Figure \ref{fig:results:Z:pTl_pTll}, we present the distribution
of the transverse momentum of the lepton, $p_{\perp,\ell}$,
alongside the transverse momentum of the system of the two leptons,
$p_{\perp,\ell\ell}$.
The transverse momentum of the leptons, $p_{\perp,\ell}$, receives small
corrections from the inclusion of higher order corrections beyond NLO
QED into the YFS formalism. Only the phenomenologically irrelevant
region of very low $p_{\perp,\ell}$ receives corrections at the permille
level at NLO EW.  Both the soft and soft-collinear approximations
agree at the permill level with NLO QED for $p_{\perp,\ell}>20$ GeV.
Correspondingly, also the dressing of the electrons has a small effect
on this distribution, with bare electrons carrying significantly less
transverse momentum than the dressed versions. The difference between
lepton species is marginal, up to about 5\% at very low $p_{\perp,\ell}$
and above the Jacobi peak.

In contrast, the transverse momentum of the system of leptons,
$p_{\perp,\ell\ell}$, shows significantly larger effects. Of course this
distribution is not defined at LO and correspondingly it is very
sensitive to the modelling of photon radiation.
This can be appreciated when comparing the NLO QED prediction with the
soft and soft-collinear approximations.  Only at small $p_{\perp,\ell\ell}$
the approximations agree.  In this observable also the inclusion of
NLO EW effects shows a significant impact, with differences reaching
up to 5\%. 
The NNLO QED effects provide a competing effect to the NLO
EW corrections, lifting the distributions by about 2\% across the
entire distribution.
The effects of the dressing on the distribution is unsurprisingly very
large as well. Bare electrons show significantly more events at
non-vanishing values of $p_{\perp,\ell\ell}$, while a different dressing
parameter leads to an almost flat decrease across the spectrum. The
comparison of the different lepton species shows that the muons again
radiate a lot more, with up to 75\% more events at medium
$p_{\perp,\ell\ell}$. $\tau$'s in comparison show a reduction in the number
of events at large $p_{\perp,\ell\ell}$ of up to 50\%.

Finally, in Figure \ref{fig:results:Z:SumEg_Phistar}, we show the
distribution of the sum of the photon energies in the decay rest
frame, $\sum_{n_{\gamma}} E_{\gamma}$, and the distribution of
the so-called $\phi_{\eta}^{\ast}$-variable.
The sum of the photon energies is largely correlated with the
$p_{\perp,\ell\ell}$, as discussed before.  This distribution shows a
distinct edge at about half the $Z$-boson mass, which is being washed
out by multiple radiation. The kinematics of the decay restrict the
energy of a single radiated photon to be smaller than
%
$E_{\gamma, \mr{max}}^1 = \frac{\hat{s}-4m_\ell^2}{2\sqrt{\hat{s}}}$,
%
which is roughly equal to half the boson mass near the resonance.  For
an event to have a total photon energy beyond this edge, two hard
photons need to recoil at least partly against each other. The region
above the kinematical edge is then only described approximately, as
long as no NNLO corrections are considered.
The NLO EW prediction mildly increases the number of events without
photon radiation, leading to a decrease at the kinematic edge of about
3\%. The NNLO QED corrections again provide a competing effect,
leading to a difference of about 1\% to the NLO QED predictions near
the edge. Beyond it, the NNLO QED corrections show a significant
departure from the shape of the previous predictions as this region is
for the first time described correctly at fixed-order.
The behaviour of different dressings and lepton species is very
similar to the case of the $p_{\perp,\ell\ell}$. The bare electrons show a
significantly larger number of hard photons, while another dressing
only leads to an approximately flat decrease. Muonic decays contain a
larger number of events with hard photons, while $\tau$'s radiate
significantly less.

The $\phi_{\eta}^{\ast}$-variable \cite{Banfi:2010cf} can be seen as
an alternative to $p_{\perp,\ell\ell}$, with the aim of being easier
measurable. It is defined purely in terms of lepton directions as:
\begin{equation} 
\phi_{\eta}^{\ast} = \tan\left(\frac{\phi_{\mr{acop}}}{2}\right)\sin\left(\theta_{\eta}^{\ast}\right),
\end{equation}
where the acoplanarity angle $\phi_{\mr{acop}}$ is defined in terms of
the difference in azimuthal angles $\Delta \phi$ between the two
leptons as $\phi_{\mr{acop}} = \pi - \Delta \phi$, and
$\theta_{\eta}^{\ast} = \tanh\left(\frac{\eta^- - \eta^+}{2}\right)$
in terms of the lepton pseudorapidities $\eta^i$. In this observable,
the soft region corresponds to the region of low $\phi_{\eta}^{\ast}$.
In comparison to the NLO QED predictions, the soft approximation
predicts too many events with low $\phi_{\eta}^{\ast}$, the difference
quickly reaches beyond 10\%. The soft-collinear approximation shows
the opposite behaviour, predicting too many events with large
$\phi_{\eta}^{\ast}$. The NLO EW prediction provide corrections of a
few percent, while the NNLO QED corrections compensate the NLO EW
corrections almost completely.
The dressing shows effects of up to 25\% at medium value of
$\phi_{\eta}^{\ast}$.

\begin{figure}[t] \centering
\includegraphics[scale=\figscale]{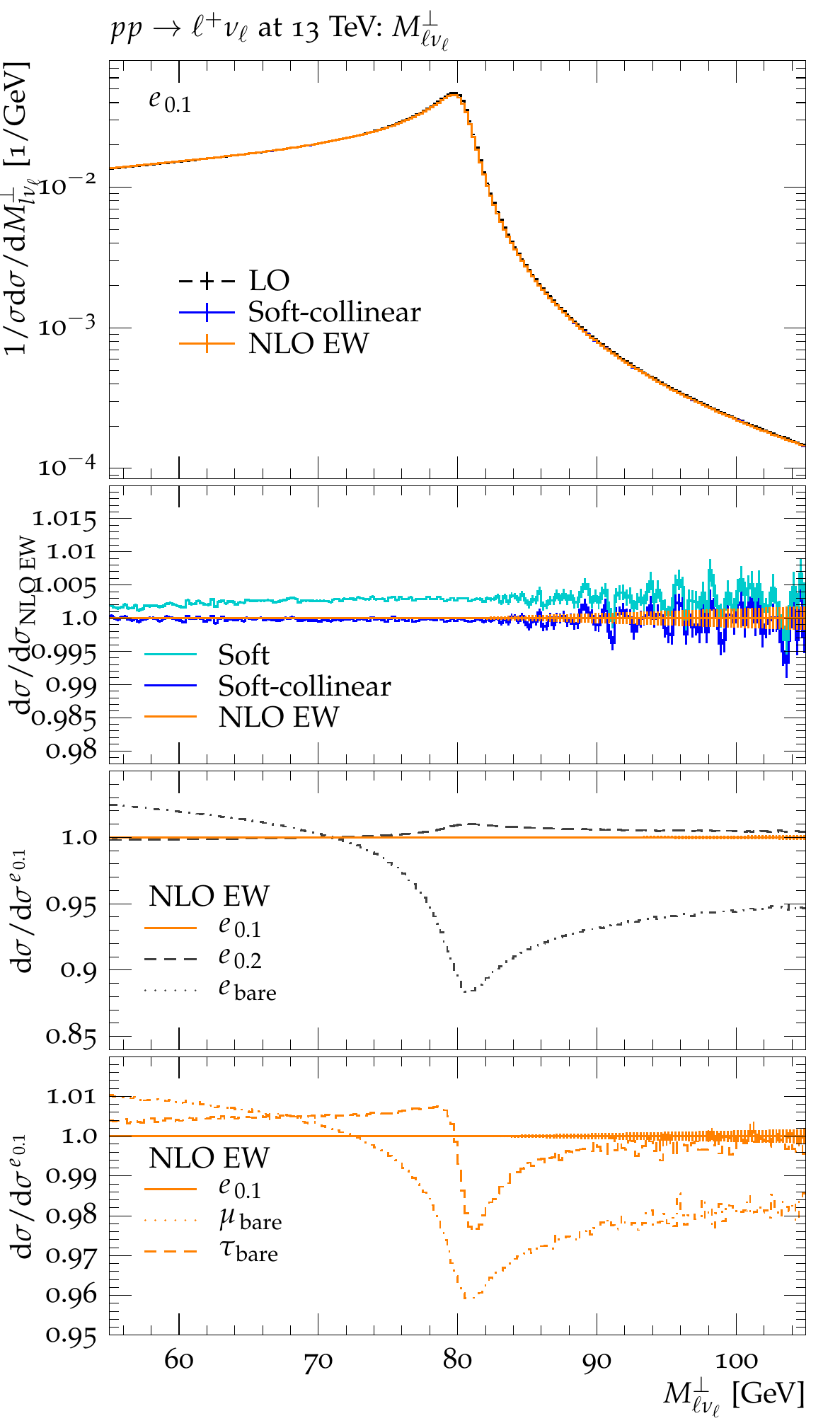}
\includegraphics[scale=\figscale]{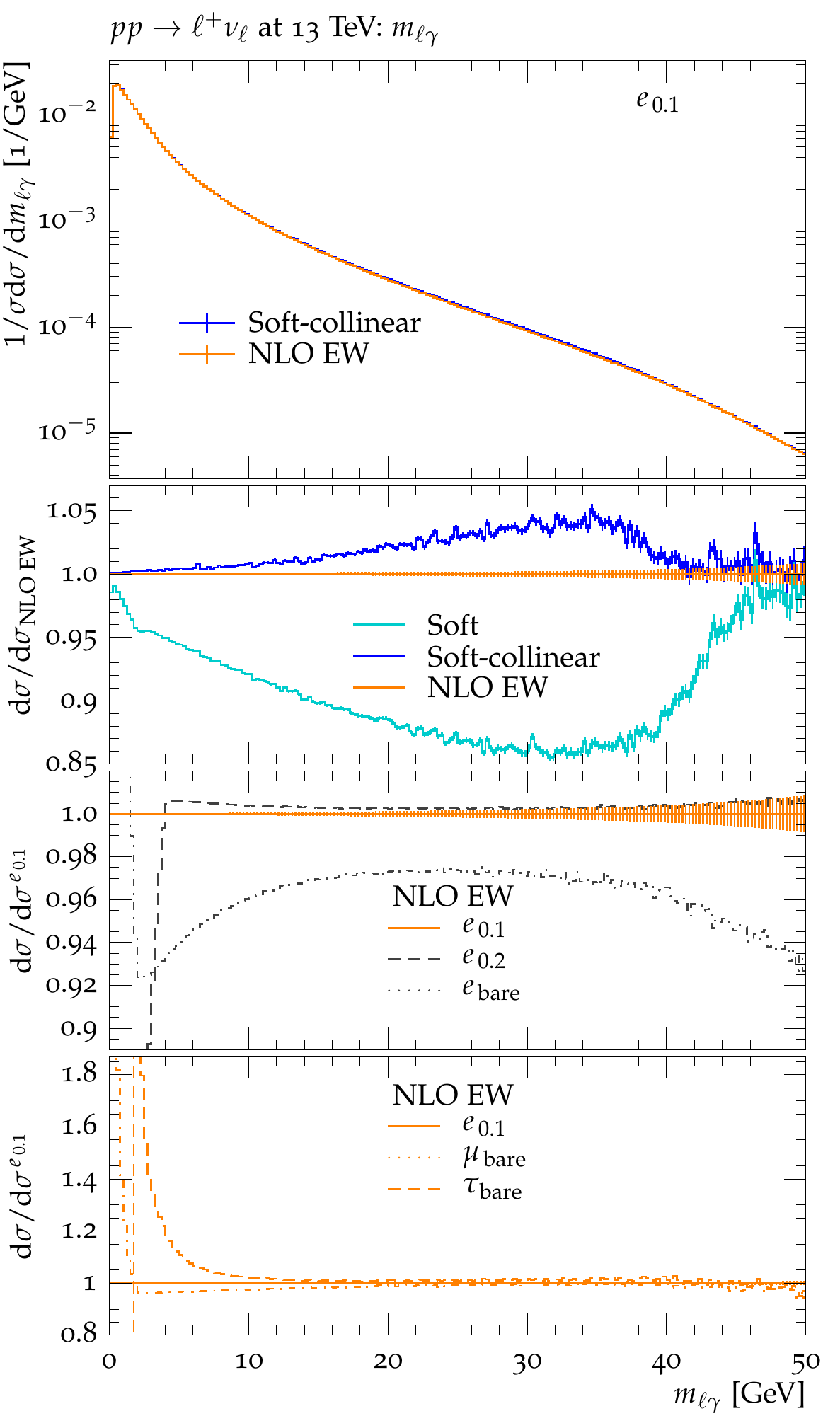}
\caption{Transverse mass of the lepton-neutrino system
  $M^{\perp}_{\ell\nu}$ (left) and the invariant mass of the system of
  the charged lepton and the nearest photon, $m_{\ell\gamma}$ (right) in
  $pp \to W^+ \to \ell^+ \nu_\ell$.  Nominal predictions are shown for $pp
  \to W^+\to e^+ v_e$ at LO, in soft-collinear NLO approximation and
  at NLO EW, where electrons are always dressed with collinear photons
  within $\Delta R=0.1$. The ratio plots highlight the effect of the
  considered higher-order corrections and the effect due to different
  photon dressing or lepton identity. See text for details.  }
\label{fig:results:W:mTlnu_mlg}
\end{figure}

\subsection{Charged Drell-Yan lepton-neutrino pair production}
\label{sec:results:W}

\begin{figure}[t] \centering
\includegraphics[scale=\figscale]{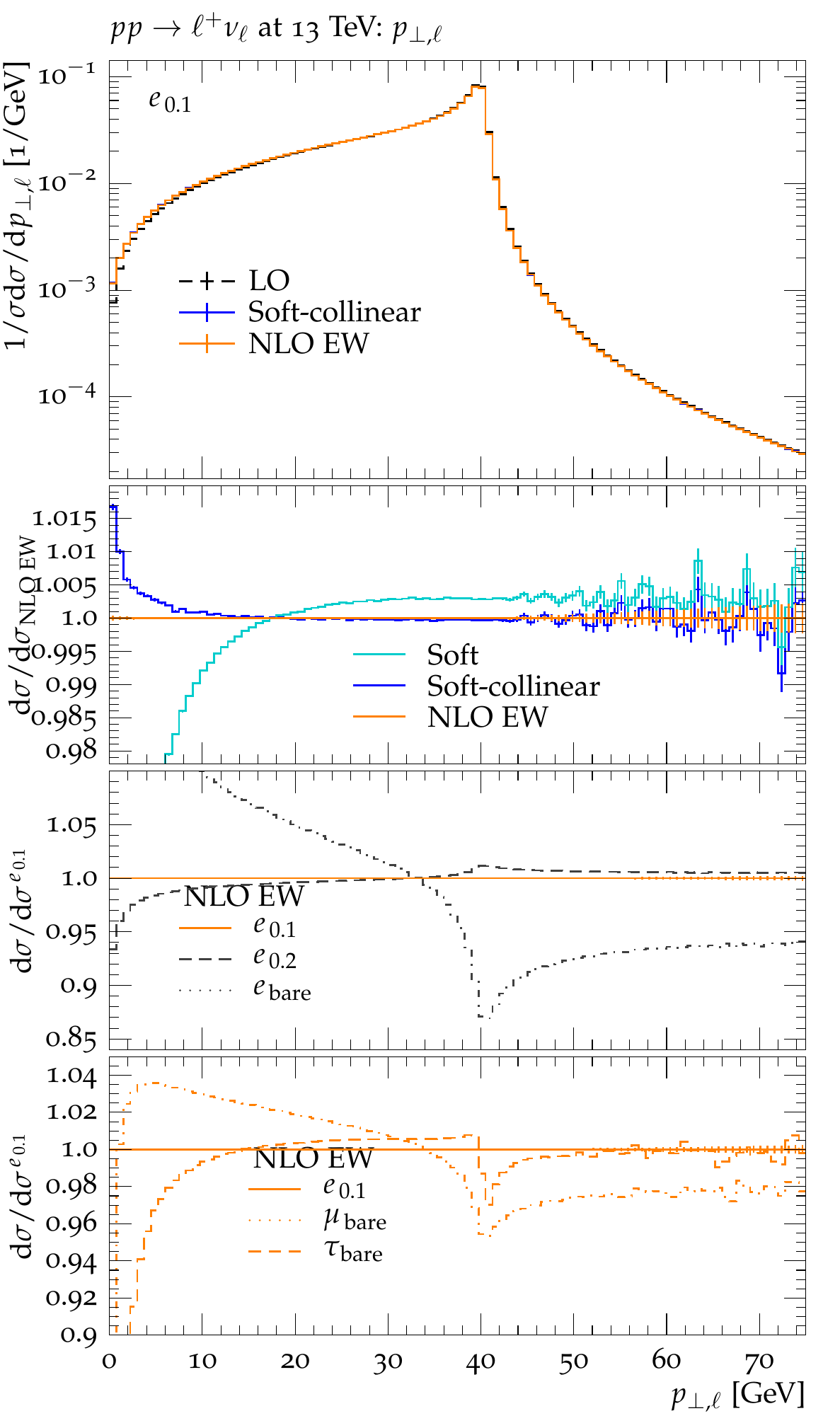}
\includegraphics[scale=\figscale]{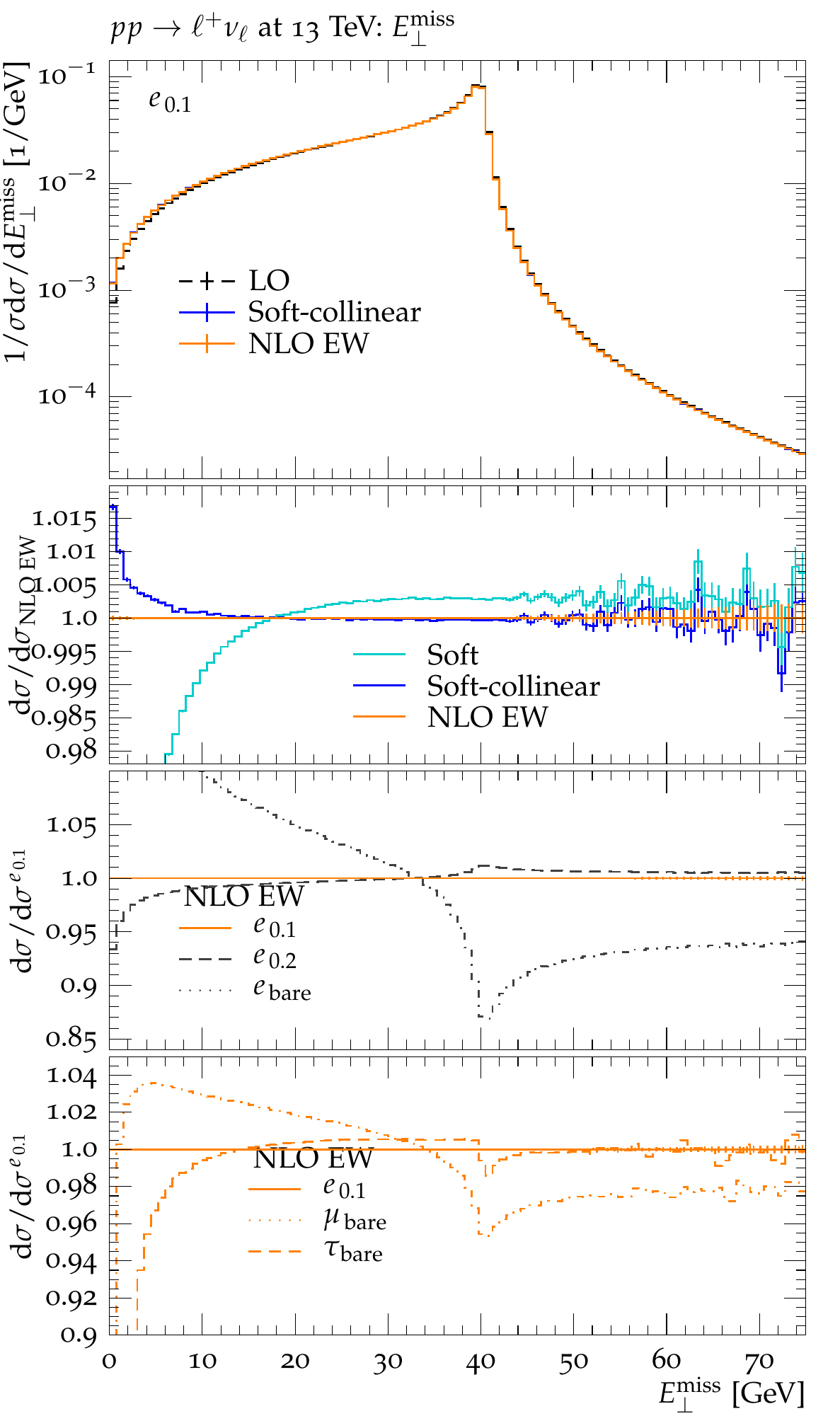}
\caption{Plots of the transverse momentum of the charged leptons,
  $p_{\perp,\ell}$, on the left and the missing transverse E,
  $E_{\perp}^{\mr{miss}}$, on the right. Predictions and labels
  as in Fig.~\ref{fig:results:W:mTlnu_mlg}. }
\label{fig:results:W:pTl_ETmiss}
\end{figure}

\begin{figure}[t] \centering
\includegraphics[scale=\figscale]{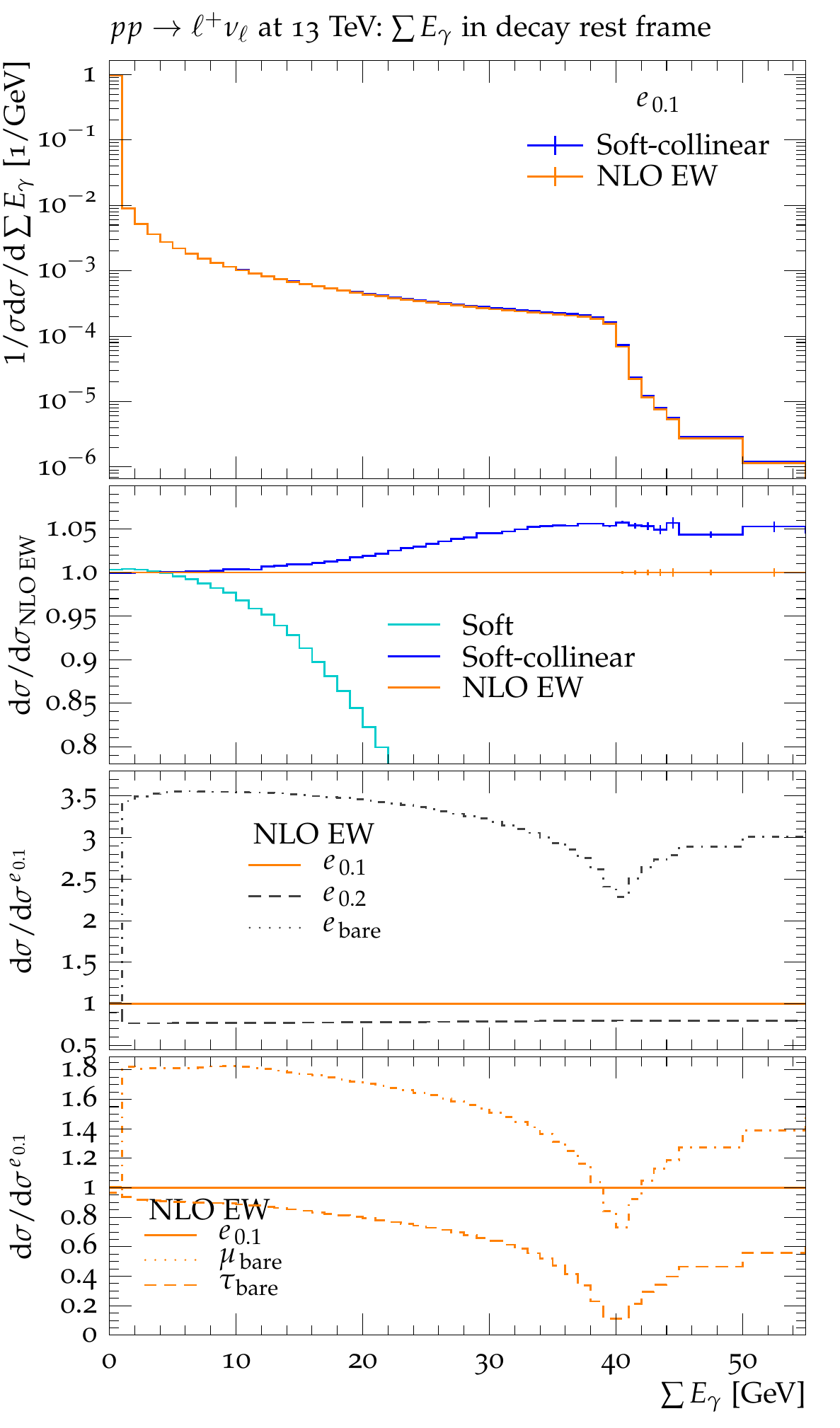}
\includegraphics[scale=\figscale]{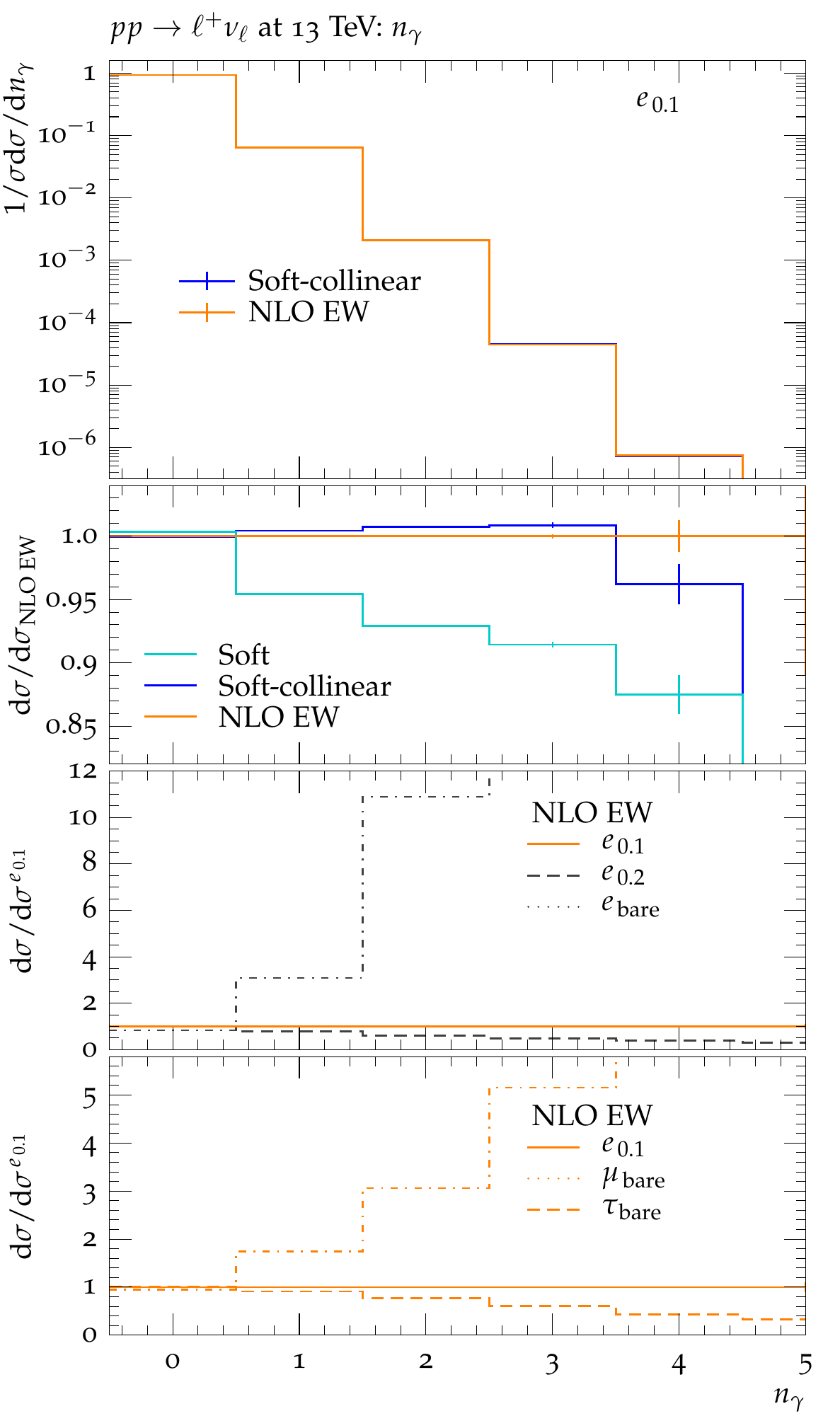}
\caption{Plots of the sum of the photon energies in the decay rest
  frame, $\sum_{n_{\gamma}} E_{\gamma}$, on the left and the number of
  photons with $E_{\gamma} > 0.1 \GeV$, $n_{\gamma}$, on the right. Predictions and labels
  as in Fig.~\ref{fig:results:W:mTlnu_mlg}.}
\label{fig:results:W:SumEg_ngamma}
\end{figure}

In
Figs. \ref{fig:results:W:mTlnu_mlg}-\ref{fig:results:W:SumEg_ngamma},
observables crucial for the study of charged-current Drell-Yan
dilepton production are investigated.  We present results for the
decay $W^+ \to \ell^+\nu_\ell$, as the charge conjugate case behaves
practically identically.  All figures are similar to the
neutral-current case presented in Section
\ref{sec:results:Z}. However, here the best prediction is of NLO EW,
as pure QED corrections cannot be defined in a gauge-invariant way.
As before all nominal predictions correspond to dressed electrons.


In Figure \ref{fig:results:W:mTlnu_mlg}, we start with the transverse mass
of the lepton neutrino system, $M^{\perp}_{\ell\nu}$, and the invariant mass of
the charged lepton and the nearest photon, $m_{\ell\gamma}$.
The $M^{\perp}_{\ell\nu}$ observable is barely affected by the NLO EW
corrections. In fact the soft-collinear approximation agrees with NLO
EW at the permill level.
The dressing of the electrons has a rather large impact, with
differences with respect to a bare treatment reaching up to 10\% at
the edge.  A slight shift of the edge is observed when comparing
different lepton species with one another, affecting the distribution
to up to a few \%.

The invariant mass of the charged lepton and the closest photon,
$m_{\ell\gamma}$ shows significantly larger corrections. Compared to the
NLO EW corrections, the soft approximation predicts a spectrum that is
too soft, while the soft-collinear approximation produces up to 5\%
more events with large $m_{\ell\gamma}$.
Bare electrons have a lot more events at low $m_{\ell\gamma}$ coming
from those photons that have not been clustered in comparison to the
dressed cases. On the other hand, those electrons dressed with $\Delta
R = 0.2$ have a reduced number of events at low $m_{\ell\gamma}$. The
comparison between lepton species shows significant differences close
to low $m_{\ell\gamma}$, illustrating the differing size of the dead
cone.

In Figure \ref{fig:results:W:pTl_ETmiss}, we show the transverse
momentum of the charged lepton, $p_{\perp,\ell}$, alongside the missing
transverse energy, $E_{\perp}^{\mr{miss}}$. The latter corresponds in
our simple setup to the transverse energy that the neutrino carries
away.
Both distributions are related and indeed they behave very similarly.
As in the neutral-current case, the transverse momentum of the charged
lepton is barely affected by NLO EW corrections, with corrections only
becoming appreciable for very low values of $p_{\perp,\ell}$.
The dressing affects the distributions by up to about 10\% in the peak
region, while different lepton species differ by up to 4\% in the peak
region and at low $p_{\perp,\ell}$.

In Figure \ref{fig:results:W:SumEg_ngamma}, we present the sum of
photon energies in the decay rest frame, $\sum_{n_{\gamma}}
E_{\gamma}$, and the number of photons with energy $E_{\gamma} > 0.1
\GeV$, $n_{\gamma}$.
The sum of photon energies shows a kinematic edge just as in the
neutral current case.  While the soft approximation predicts too soft
a spectrum of photon energies, the soft-collinear approximation does a
much better job in $W$-decays as the effects coming from NLO EW
corrections reach at most 5\% at the kinematic edge.  The reason for
this behaviour can be read from the distribution of the
$n_{\gamma}$. The soft approximation is shown to produce too few
photons, while the soft-collinear approximation predicts more events
with 1-3 photons.
Analyses using bare electrons show a significantly larger number of
photons, with already 4 times more events with 1 photon. At the same
time, for $\Delta R = 0.2$ electrons, the number of photons is suppressed
significantly. A similar picture presents itself when comparing lepton
species. Muonic decays contain significantly more photons, while
decays into $\tau$'s end up with a lot less events with at least one
photon.

As a noteworthy observation we want to point out a difference between
neutral-current and charged-current processes: the soft-collinear
approximation is more reliable in the charged-current case.  This can
be understood from the fact that here collinear radiation
predominantly originates from just one particle, the lepton, rather
than two competing particles as in the $Z$-boson case. Any error due
to the missing interference contributions in the soft-collinear
approximation is thus significantly diminished.

\subsection{Leptonic Higgs-boson decays}
\label{sec:results:h}

\begin{figure}[t] \centering
\includegraphics[scale=\figscale]{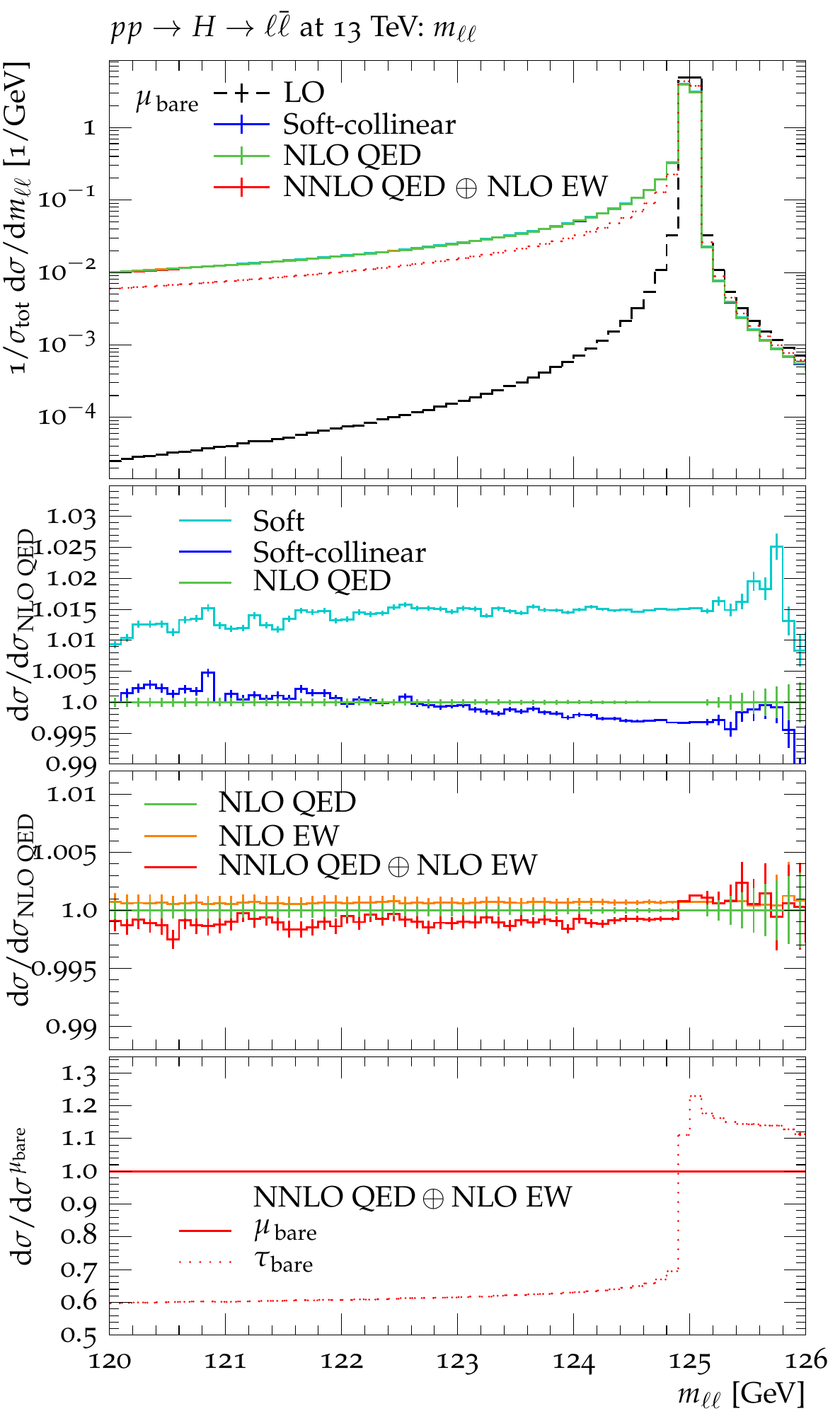}
\includegraphics[scale=\figscale]{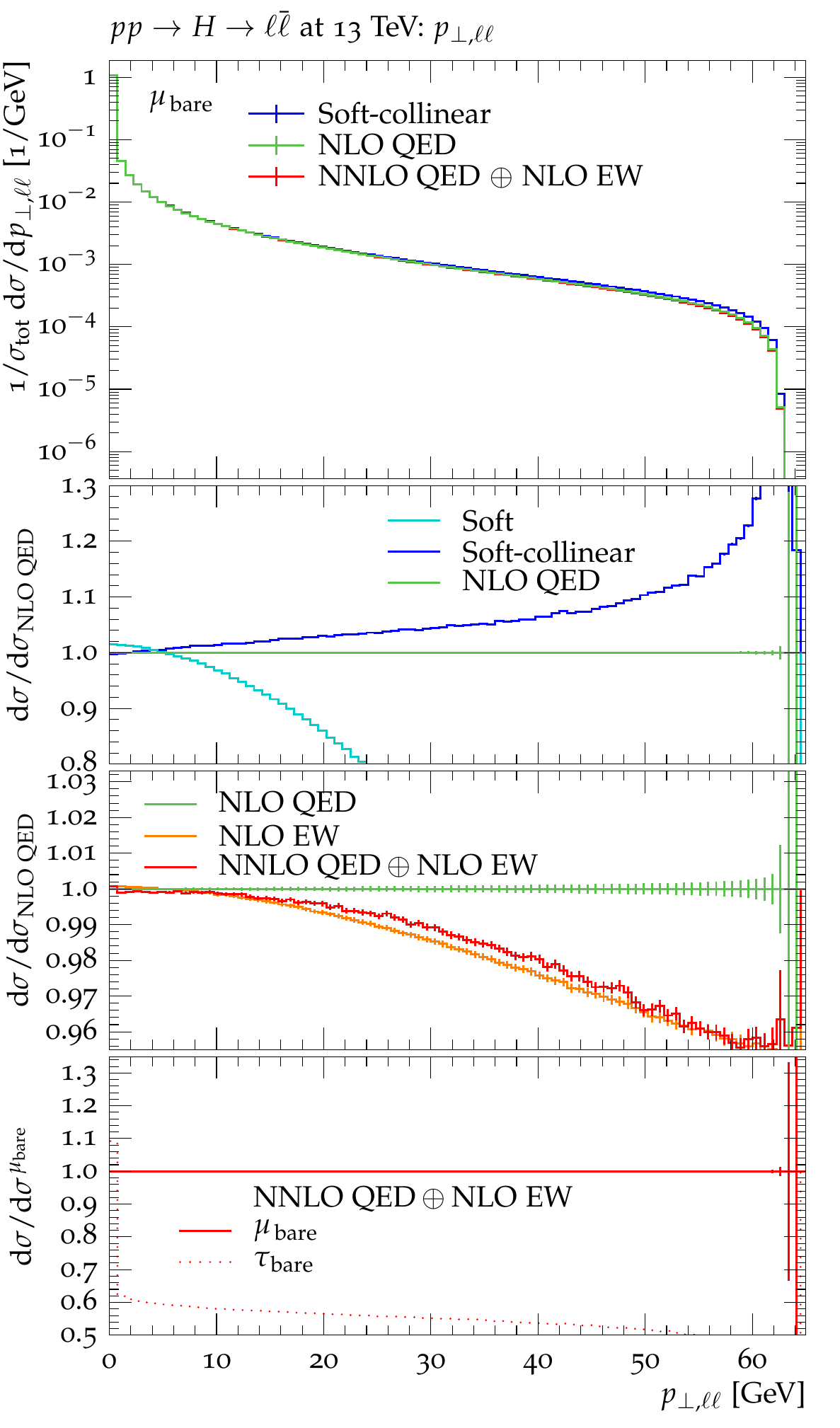}
\caption{Plots of the invariant mass of the two decay leptons,
  $m_{\ell\ell}$, on the left and the transverse momentum of the system of
  the two leptons, $p_{\perp,\ell\ell}$, on the right in the process $pp\to
  H \to \ell^+\ell^-$. Nominal predictions are shown for $pp \to H\to \mu^+
  \mu^-$ at LO, in soft-collinear NLO approximation, at NLO QED and at
  NNLO QED $\oplus$ NLO EW. The ratio plots highlight the effect of
  the considered higher-order corrections and lepton identity. See
  text for details.}
\label{fig:results:H:mll_pTll}
\end{figure}

Finally we highlight the effect of higher-order corrections in photon
radiation off leptonic Higgs decays. Numerical results are shown in
Fig.\  \ref{fig:results:H:mll_pTll}, where the nominal distribution
corresponds to $H\to \mu^+ \mu^-$ with bare muons. Here we focus on the
dilepton invariant mass $m_{\ell\ell}$ and $p_{\perp,\ell\ell}$ recoil. As for
neutral-current Drell-Yan we consider higher-order corrections at the
level of soft and soft-collinear approximations, full NLO QED, NLO EW
and also combining NLO EW with NNLO QED. The LO prediction clearly
fails to describe the invariant mass distribution. Yet, the soft and
soft-collinear approximations provide a quite reliable description
with corrections smaller than 1-2\% with respect to full NLO QED. The
weak corrections are slightly larger compared to the neutral-current
Drell-Yan case, still they alter the invariant mass distribution only
at the permille level and are overcompensated by NNLO QED effects of
the same order.  As mentioned in Section~\ref{sec:results_setup} we
are unable to resolve the sharp mass peak of the Higgs-boson with the
lowest energy photons we generate. However, investigating the low
energy tail of the invariant mass distribution, we observe that the
NLO QED corrections provide a mostly flat contribution in the peak
region.
Comparing decays into bare muons with decays into bare $\tau$'s, we
can appreciate a significantly smaller sensitivity of the $\tau$
distribution to QED radiation.

The distribution of the transverse momentum of the di-lepton system
shows similar effects as in the case of the $Z$-boson decay. The soft
approximation predicts a distribution that is far too soft, while the
soft-collinear approximation predicts too many events with large
$p_{\perp,\ell\ell}$. The NLO EW corrections increase the number of events
by about a permille at low $p_{\perp,\ell\ell}$, and decrease them at high
values up to about 5\%. The NNLO QED corrections in this case do not
provide a large competing effect, and the NNLO QED $\oplus$ NLO EW
prediction agrees with the NLO EW one at the permille level.
Decays into $\tau$'s show about 40\% less events with non-vanishing
$p_{\perp,\ell\ell}$, the effect being close to constant across the entire
distribution.

\section{Conclusions and outlook}
\label{sec:conclusions}

In this paper, we have presented an implementation of NLO EW and NNLO
QED corrections to the decays of weak gauge and Higgs bosons within the YFS
formalism. For this purpose, we extended \Sherpa's module \Photons to
include the relevant matrix elements, renormalized in the on-shell
scheme, and subtractions needed within this formalism. In our
numerical results we find that observables relating only to the leptons in
the process are only marginally affected by the corrections, up to the
level of a few permil. In particular, the peak of the invariant mass
distributions is practically not affected. Distributions that relate
to the energies of the generated photons themselves, or can be related
to them, such as the transverse momentum of the pair of the leptons
$p_{\perp,\ell\ell}$, naturally receive larger corrections. The electroweak
corrections increase the likelihood of hard photon radiation by up to
2-3\% for very hard radiation. The NNLO QED corrections compete with
these corrections by reducing the likelihood of hard radiation, albeit
to a smaller extent. At the same time, some regions of phase space are
only described at leading order in $\alpha$ upon the inclusion of the
double real radiation, such that in these regions the corrections can
be significantly larger. Examples for such regions are those where
the sum of the photon energies exceeds half the boson mass or regions
of large $\phi^{\ast}$. Angular distributions of the photons are not
affected by higher order contributions confirming the general
radiation pattern of QED radiation.  The results give us confidence
that the inclusion of the full EW corrections to particle decays
within the YFS formalism in \Sherpa are sufficient to achieve precise
results for most leptonic observables.  Beyond the corrections
investigated in this work, it will be interesting to consider the YFS
formalism also including initial state effects and the matching to
NLO EW corrections to the hard production process, see~\cite{Kallweit:2017khh}.

The implemented NNLO QED and NLO EW corrections provide high precision
also in extreme phase space regions and can be seamlessly added to
standard precision QCD simulations.  This provides an important
theoretical input to future precision determinations of fundamental
parameters of the EW theory at hadron colliders and beyond.

\subsection*{Acknowledgements}

We would like to thank our colleagues from the \Sherpa and \OpenLoops
collaborations for discussions.  We are indebted to D.~Wackeroth and
A.~Vicini for support with \WZGRAD and for clarification of sometimes
subtle issues in the calculational framework.
This work was financially supported by the European Commission under
Grant Agreements PITN-GA-2012-315877 (“MCnet”) and PITN-GA-2012-316704
(“HiggsTools”), and by the ERC Advanced Grant 340983 (“MC@NNLO”).

\appendix
\label{sec:app}
\section{Momentum mappings}
\label{sec:impl:MomMapping}

For the purposes of event generation, we need to define the momenta
that are used in the master formula \EqRef{sec:impl:eq:master}. We
will refer to the momenta used in the leading order matrix element,
$\tilde{\beta}_0^0$, as the ``undressed'' momenta, {\it i.e.} the
momenta before the event is dressed with photons. The undressed
momenta are labelled through $q_i^{\mu}$, and we define as
\begin{equation}
  Q_{N/C}^{\mu} = \sum_{i\in N/C FS} q_i^{\mu}\,
\end{equation}
the sums of the final state neutral and charged momenta. After the
generation of the additional photon momenta, the undressed momenta
have to be mapped to a set of ``dressed'' momenta to account for
momentum conservation. The dressed momenta are labelled through
$p_i^{\mu}$ and we define the sums of the neutral and charged final
state particles in the same way as for the undressed momenta:
\begin{equation}
  P_{N/C}^{\mu} = \sum_{i\in N/C FS} p_i^{\mu}\,.
\end{equation}
In a similar manner, we define the sum of the photon momenta as
\begin{equation}
  K^{\mu} = \sum_{i = 1}^{n_R} k_i^{\mu}\,.
\end{equation}
The mappings relevant for particle decays
of both uncharged and charged initial particles have been outlined in
section 3.3 of \cite{Schonherr:2008av}, but we will repeat them here
for the benefit of the interested reader.  The only condition the mapping
has to meet is that in the limit of $K\to 0$,
the underlying momenta of the undressed $n$-parton phase space have to
be recovered exactly. QED provides no guiding principle which particle
should be taken to balance the momenta of the generated photons. It is
therefore sensible to treat all the final state momenta fully
democratically and let them all take the recoil. Considering all
particles in the rest frame of the multipole responsible for the
radiation, this can be achieved by scaling the three-momenta of all
final state particles by a common factor $u$, distributing the photon
momenta across, and finally enforcing momentum conservation and
on-shell conditions.

\subsection{Neutral initial states}

For a neutral particle of mass $m$ decaying into charged particles,
such as is the case of a $Z$- or a Higgs-boson, the
above procedure fixes the mapping to a rescaling of all final state momenta,
balancing the photonic momentum by moving the frame of the multipole.

We start with the undressed momenta in the multipole rest frame
\begin{align}
  \begin{split} 
    q^{\mu} &= \left(\sqrt{m^2+\vec{Q}_N^2},\vec{Q}_N\right), \\
    Q_C^{\mu} &= \left(Q_C^0,\vec{Q}_C = \vec{0}\right), \\
    Q_N^{\mu} &= \left(Q_N^0,\vec{Q}_N\right)\,.
  \end{split}
\end{align}
The outlined procedure maps these momenta onto the final state momenta
$P_C$ and $P_N$:
\begin{align}
\begin{split} 
{q'}^{\mu} &= \left(\sqrt{m^2+\left(u\vec{Q}_N+\vec{K}\right)^2},u\vec{Q}_N+\vec{K}\right), \\ 
P_C^{\mu} &= \left(P_C^0,u\vec{Q}_C = \vec{0}\right), \\ 
P_N^{\mu} &= \left(P_N^0,u\vec{Q}_N\right), \\
K^{\mu} &= \left(K^0,\vec{K}\right)\,.
\end{split}
\end{align}
We can rewrite the three momentum of the initial state as
$u\vec{Q}_N+\vec{K} = u\vec{q}+\vec{K}$ showing that the two vectors
$q$ and $q'$ are the same vector in different frames. All momenta now
reside in the rest frame of the dressed multipole. We can then
determine the scaling parameter $u$ from energy conservation:
\begin{equation} 
0 = \sqrt{m^2+\left(u\vec{Q}_N+\vec{K}\right)^2} -
\sum_C \sqrt{m_i^2+u\vec{q_i}^2} - \sum_N \sqrt{m_i^2+u\vec{q_i}^2} -
K^0\,.
\end{equation}

\subsection{Charged initial states}

For a charged particle of mass $m$ decaying into a charged particle
and a number of neutral particles, such as is the case of a  $W$-boson, 
we require a different approach. In order to remain in
the rest frame of the dressed multipole, we cannot accomodate the
photon momenta purely in the initial state.

Again, we start with the undressed momenta in the multipole rest
frame:
\begin{align}
\begin{split} 
q^{\mu} &= \left(\sqrt{m^2+\vec{Q}_C^2},-\vec{Q}_C\right), \\ 
Q_C^{\mu} &= \left(Q_C^0,\vec{Q}_C\right), \\
Q_N^{\mu} &= \left(Q_N^0,\vec{Q}_N = -2\vec{Q}_C\right)\,.
\end{split}
\end{align}
In the most democratic approach, the photon momenta are accomodated
equally by all particles in the final state and the undressed momenta
will be mapped onto:
\begin{align}
\begin{split} 
{q'}^{\mu} &= \left(\sqrt{m^2+\left(-u\vec{Q}_C+n_C\vec{\kappa}\right)^2},-u\vec{Q}_C+n_C\vec{\kappa}\right), \\
P_C^{\mu} &= \left(P_C^0,u\vec{Q}_C - n_C\vec{\kappa} \right), \\
P_N^{\mu} &= \left(P_N^0,u\vec{Q}_N - n_N\vec{\kappa}\right), \\
K^{\mu} &= \left(K^0,\vec{K}\right).
\end{split}
\end{align}
All momenta now reside in the rest frame of the dressed multipole. The
$n_C$ and $n_N$ denote the number of charged and neutral final state
particles, and $\vec{\kappa}$ is defined as:
\begin{equation}
  \vec{\kappa} = \frac{1}{2n_C+n_N} \vec{K}.
\end{equation}
One can however also choose to let only the charged particles or only
the neutral particles in the process accomodate the photon momenta, in
which case $\vec{\kappa} = \vec{K}/(2n_C)$ or $\vec{\kappa} =
\vec{K}/(n_N)$, respectively, and corresponding terms in the mapping
vanish. The default option in \Photons, and the one that we employ for the results presented this paper, is the choice of letting only
the neutral particles take the recoil.

Again, the scaling parameter $u$ can be determined from energy
conservation:
\begin{equation} 0 =
\sqrt{m^2+\left(-u\vec{Q}_C+n_C\vec{\kappa}\right)^2} - \sum_C
\sqrt{m_i^2+\left(u\vec{q_i}-\vec{\kappa}\right)^2} - \sum_N
\sqrt{m_i^2+\left(u\vec{q_i}-\vec{\kappa}\right)^2} - K^0.
\end{equation}

\subsection{Momenta in higher order corrections}

Having discussed the momentum mappings necessary to map from undressed
to dressed momenta, it is worth briefly discussing which set of
momenta is to be used in each component of \EqRef{sec:impl:eq:master}.

Every part of this formula apart from the correction factor, $\mc{C}$,
is calculated using the undressed momenta $q_i$, with the Jacobean
$\mc{J}$ accounting for the mapping from undressed to dressed
momenta. This in particular includes the factors $\tilde{S}$ that
implement the soft approximation to the real matrix elements.

The correction factor $\mc{C}$ amounts to a reweighting of the 
YFS approximation to the required perturbative order. 
Practically, for the real matrix element corrections, the
eikonal factors $\tilde{S}$ have to be cancelled out. Thus, the
eikonals in the denominators in \EqRef{sec:impl:eq:C} have to be
calculated using the undressed momenta.

All matrix elements containing no additional photon,
$\tilde{\beta}_0^i$, are calculated in the $n$-particle Born phase
space, {\it i.e.} using the undressed momenta. The terms describing real
matrix element corrections are then calculated in the phase space
appropriate to the number of photons they contain: In the
$n+1$-particle phase space for the single real matrix elements
$\tilde{\beta}_1^i$, in an $n+2$-particle phase space for double real
matrix elements, $\tilde{\beta}_2^i$, and so on. In order to define
the momenta in these phase spaces, we repeat the mapping procedure
described previously, but now only taking into account the photons
that are taken to be hard in the matrix element correction. This
procedure is repeated for every photon or set of photons that have
been created. For the single real matrix elements, this means there
are in total $n_{\gamma}$ calls to the mapping and the matrix
elements, while for the double real matrix elements, there are
$n_{\gamma}(n_{\gamma}-1)/2$ calls to the mapping and the matrix
element.

\section{NLO EW form factors and counterterms}
\label{sec:app:NLO_EW}

For completeness in this section we collect the electroweak vertex form factors and
counterterms required for setting up the NLO electroweak corrections
to $\tilde{\beta}_0^1$. We use the vertex form factors found in
\cite{Bardin:1999ak} and the counterterms in the on-shell
renormalization scheme from \cite{Denner:1991kt}. The vertex form
factors retain the full dependence on the lepton masses only in the
QED corrections, while the purely weak contributions are calculated in
the massless limit.
In order to find the pure NLO QED corrections, out of the form factors
we need to only include the QED form factors. In the counterterms, we
only need to include the photonic corrections to the wavefunction
renormalization. The tree-level couplings are formally purely weak
couplings and do not need to be renormalized in this case. This
procedure amounts to the full separation of the $U(1)$- from the
$SU(2)$-theory. Such a procedure is only possible for the decays of
electroweak particles that do not couple to the photon, in our case
the $Z$- and Higgs bosons. For the charged $W$-boson, such a
separation does not yield a gauge-invariant subset of contributions
and we can only calculate matrix elements in the full electroweak
theory.
Here we note, that shifts due to a different metric signature in \Sherpa and \cite{Denner:1991kt}
versus  \cite{Bardin:1999ak} have been considered.
All results presented here are in Feynman gauge. We call the left- and
right-handed tree-level couplings $c_L$ and $c_R$ and introduce $g_L =
c_L \frac{s_Wc_W}{ie}$, $g_R = c_R \frac{s_Wc_W}{ie}$ for convenience.
We further use the vector coupling $v_f = (g_L+g_R)$ and the axial
coupling $a_f = (g_L-g_R)$. Any quantity denoted as $x_{f'}$ refers to
the iso-spin partner of the fermion~$f$.

\subsection{\texorpdfstring{$Z\to f\bar{f}$}{Z->ffbar}}

The QED corrections to this vertex are given by:%
\begin{align}
\begin{split}
V_{\mu}^{\mr{QED}} =& \frac{\alpha}{4\pi} \frac{e}{2s_Wc_W}Q_f^2 \bigg[i\gamma_{\mu}\left(v_f-a_f\gamma_5\right)F_{Aa}(s) - i\gamma_{\mu}a_f\gamma_5F_A^{(1)}(s) \\
&+ v_f\left(p_f-p_{\bar{f}}\right)_{\mu}F_V^{(2)}(s) - a_f\gamma_5\left(p_f+p_{\bar{f}}\right)_{\mu}F_A^{(3)}(s)\bigg] \\
=& \frac{\alpha}{4\pi} \frac{1}{s_Wc_W}Q_f^2 \bigg[\gamma_{\mu}\left(c_L P_L+c_RP_R\right)F_{Aa}(s) -ie\frac{I_f}{2}\gamma_{\mu}\left(P_R-P_L\right)F_A^{(1)}(s) \\
&+ e\frac{I_f-2s_W^2Q_f}{2}\left(p_f-p_{\bar{f}}\right)_{\mu}\left(P_R+P_L\right)F_V^{(2)}(s)\\
&- e\frac{I_f}{2}\left(P_R-P_L\right)\left(p_f+p_{\bar{f}}\right)_{\mu}F_A^{(3)}(s)\bigg].
\end{split}
\end{align}
In the massless limit, only the structure proportional to
$\gamma_{\mu}\left(c_L P_L+c_RP_R\right)$ contributes. The form factor
$F_{Aa}(s)$ is given by:%
\begin{align}
\begin{split}
F_{Aa}(s) =& -2\left(s-2m_f^2\right)C_0\left(m_f^2,m_f^2,s,m_f^2,0,m_f^2\right) 
- 3B_0\left(s,m_f^2,m_f^2\right) + 4B_0\left(m_f^2,m_f^2,0\right) -2.
\end{split}
\end{align}
The other form factors are all proportional to the fermion mass and
read%
\begin{align}
F_V^{(2)}(s) =& \frac{2m_f}{4m_f^2-s}\left[B_0\left(s,m_f^2,m_f^2\right)-B_0\left(m_f^2,m_f^2,0\right)\right], \\
F_A^{(1)}(s) =& -\frac{8m_f^2}{4m_f^2-s}\left[B_0\left(s,m_f^2,0\right)-B_0\left(m_f^2,m_f^2,0\right)\right], \\
F_A^{(3)}(s) =& \frac{m_f}{s\left(4m_f^2-s\right)}\left[\frac{4m_f^2-3s}{2}\left(B_0\left(s,m_f^2,0\right)-B_0\left(m_f^2,m_f^2,0\right)\right)+4m_f^2-s\right].
\end{align}
The effect of abelian $Z$- and $\phi^0$-exchanges is given by:%
\begin{align}
\begin{split}
V_{\mu}^{Za} = \frac{\alpha}{4\pi} \frac{ie}{s_W^3c_W^3}\gamma_{\mu} \left(g_L^3P_L+g_R^3P_R\right) F_{Za}(s),
\end{split}
\end{align}
where
\begin{align}
\begin{split}
F_{Za} (s) =& -\frac{2M_Z^4}{s}\left(1+\frac{s}{M_Z^2}\right)^2C_0\left(0,0,s,0,M_Z^2,0\right)\\
& +B_0\left(s,0,0\right) - \left(\frac{2M_Z^2}{s}+4\right) \left[B_0\left(s,0,0\right)-B_0\left(0,0,M_Z^2\right)\right] -2 \label{eq:FZa}.
\end{split}
\end{align}

For the diagrams involving $W$ bosons (and the associated ghosts), we introduce:%
\begin{alignat}{4}
w_f  &=& \frac{m_f^2}{M_W^2}, \quad  w_{f'} &= \frac{m_{f'}^2}{M_W^2}, \\
\beta^2 &=& 1 - w_{f'},  \quad \kappa &= -\frac{\beta^2 \left(3-\beta^2\right)}{2}\frac{M_W^2}{s}.
\end{alignat}
The effect of abelian $W$- and $\phi$-exchanges, i. e. all diagrams
not involving a three-boson vertex, is described by:%
\begin{align}
\begin{split}
V_{\mu}^{Wa} =
 \frac{\alpha}{4\pi} \frac{1}{2s_W^2}\gamma_{\mu} P_L \left[g_{L'}F_{Wa}(s) + \frac{I_{f'}}{2}\bar{F}_{Wa}(s)\right].
\end{split}
\end{align}
Note that this is purely a contribution to the left-handed part of the amplitude.
The necessary auxilliary functions are given by:
\begin{align}
\begin{split}
F_{Wa}(s) =& -\left(-2\beta^2\kappa + 3 + \beta^4 + 2\frac{s}{M_W^2}\right)M_W^2 C_0\left(0,0,s,m_{f'}^2,M_W^2,m_{f'}^2\right) \\
& +2\left(\kappa-2\right)\left[B_0\left(s,m_{f'}^2,m_{f'}^2\right)-B_0\left(0,m_{f'}^2,M_W^2\right)\right] \\
& +\frac{\left(3-\beta^2\right)}{2}B_0\left(s,m_{f'}^2,m_{f'}^2\right) - \left(2+\frac{1}{2}w_{f'}\right), \\
\bar{F}_{Wa}(s) =& w_{f'}\bigg[\left(\frac{\beta^4M_W^2}{s}+2\right)M_W^2 C_0\left(0,0,s,m_{f'}^2,M_W^2,m_{f'}^2\right) \\
& +\frac{\beta^2M_W^2}{s}\left[B_0\left(s,m_{f'}^2,m_{f'}^2\right) - B_0\left(0,m_{f'}^2,M_W^2\right)\right] \\
& -\frac{1}{2}B_0\left(s,m_{f'}^2,m_{f'}^2\right) + \frac{1}{2}\bigg].
\end{split}
\end{align}
The effect of non-abelian $W$- and $\phi$-exchanges, {\it i.e.} all the
diagrams containing a three-boson vertex, is described by:%
\begin{equation}
V_{\mu}^{Wn} = \frac{\alpha}{4\pi} \frac{ie}{s_Wc_W}\frac{c_W^2}{s_W^2} \left(-I_f\right)\gamma_{\mu} P_L \left[F_{Wn}(s) + \bar{F}_{Wn}(s)\right].
\end{equation}
Note that this is again purely a contribution to the left-handed part of the amplitude.
The necessary auxilliary functions are given by:%
\begin{align}
\begin{split}
F_{Wn}(s) =& -\left(-2\beta^2\kappa + 3 + \beta^4\right)M_W^2 C_0\left(0,0,s,M_W^2,m_{f'}^2,M_W^2\right) \\
&-2\left(\kappa-2\right)\left[B_0\left(s,M_W^2,M_W^2\right)-B_0\left(0,m_{f'}^2,M_W^2\right)\right] \\
&-\left(3+\frac{1}{2}w_{f'}\right)B_0\left(s,M_W^2,M_W^2\right) -\frac{1}{2}w_{f'}, \\
\bar{F}_{Wn}(s) =& \frac{1}{2c_W^2}w_{f'} \bigg[\left(\frac{\beta^4M_W^2}{s}-4+w_{f'}\right)M_W^2C_0\left(0,0,s,M_W^2,m_{f'}^2,M_W^2\right) \\
&-\frac{\beta^2M_W^2}{s}\left[B_0\left(s,M_W^2,M_W^2\right)-B_0\left(0,m_{f'}^2,M_W^2\right)\right] \\
&+\frac{1}{2} \left(B_0\left(s,M_W^2,M_W^2\right)+1\right)\bigg].
\end{split}
\end{align}
The counterterms for this vertex read:
\begin{align}
\delta_R &= c_R \left(1+\frac{1}{2}\delta Z_{ZZ}+\frac{1}{2} \left(\delta Z_{ii}^{f,R} + \delta Z_{ii}^{f,R,\dagger}\right)\right) + \delta c_R - \frac{1}{2} Q_f \delta Z_{AZ} \\ 
\delta_L &= c_L \left(1+\frac{1}{2}\delta Z_{ZZ}+\frac{1}{2} \left(\delta Z_{ii}^{f,L} + \delta Z_{ii}^{f,L,\dagger}\right)\right) + \delta c_L - \frac{1}{2} Q_f \delta Z_{AZ}, \label{eq:Zffcounterterm}
\end{align}
where the left- and right-handed, tree-level couplings $c_R$, $c_L$
and their counterterms $\delta c_R$, $\delta c_L$ are given by:
\begin{align}
c_R &= \frac{ie}{s_Wc_W} \left(-s_W^2 Q_f\right), \\
\delta c_R &= c_R \left(\delta Z_e + \frac{1}{c_W^2}\frac{\delta s_W}{s_W}\right), \\
c_L &= \frac{ie}{s_Wc_W} \left(I_f-s_W^2 Q_f\right), \\
\delta c_L &= \frac{ie}{s_Wc_W} I_f\left(\delta Z_e +\frac{s_W^2-c_W^2}{c_W^2} \frac{\delta s_W}{s_W}\right) + \delta c_R.
\end{align}
\subsection{\texorpdfstring{$W^-\to \ell^-\bar{\nu}_\ell$}{W->l-nu} and \texorpdfstring{$W^+\to \nu_\ell\bar{\ell}^+$}{W->l+nu}}

The decay of $W$-bosons will only be applied to $\ell\nu$ final states
within the YFS framework, whereas $q\bar{q}'$ final states will be
treated within the parton shower to allow for consistent matching with
the dominant QCD corrections. In the case of $\ell\nu$ final states,
there is no diagram for photon exchange between the final state
particles.
All the corrections to this decay are purely corrections to the
left-handed coupling (since fermion masses are neglected in these
subamplitudes).
The effect of non-abelian photon exchange is given by:
\begin{align}
V_{\mu}^{An}(s) = \frac{\alpha}{4\pi} \frac{ie}{\sqrt{2}s_W}\; 2P_L\; \mr{sgn}\left(Q_f\right) F_{An}(s).
\end{align}
The form factor is given by:%
\begin{align}
\begin{split}
F_{An}(s) =& Q_f\left[M_W^2C_0\left(m_f^2,m_{f'}^2,s,0,m_{f}^2,M_W^2\right) + B_0\left(m_f^2,m_f^2,0\right)\right] \\
&-Q_{f'}\left[M_W^2C_0\left(m_f^2,m_{f'}^2,s,M_W^2,m_{f'}^2,0\right) + B_0\left(m_{f'}^2,m_{f'}^2,0\right)\right] \\
&+\frac{Q_f-Q_{f'}}{2}\left[-\left(\frac{M_W^2}{s}+1\right)B_0\left(s,M_W^2,0\right)+\left(\frac{M_W^2}{s}+2\right)B_0\left(0,0,M_W^2\right)\right].
\end{split}
\end{align}
The effect of abelian $Z$-exchange is described by:
\begin{align}
\begin{split}
V_{\mu}^{Za}(s) 
= \frac{\alpha}{4\pi} \frac{ie}{\sqrt{2}s_W} \frac{1}{s_W^2c_W^2} \gamma_{\mu}P_Lg_Lg_{L'}F_{Za}(s).
\end{split}
\end{align}
with the function $F_{Za}(s)$ as in the decay $Z\to f\bar{f}$ (\EqRef{eq:FZa}).

The effect of non-abelian $Z$-exchange is given by:%
\begin{align}
\begin{split}
V_{\mu}^{Zn}(s) 
= \frac{\alpha}{4\pi} \frac{ie}{\sqrt{2}s_W} \frac{4}{s_W^2} \gamma_{\mu}P_L\mr{sgn}\left(Q_f\right)\left(g_L-g_{L'}\right)F_{Zn}(s).
\end{split}
\end{align}
The form factor reads:%
\begin{align}
\begin{split}
F_{Zn}(s) =& \frac{1}{2} \bigg\{\left[\left(\frac{M_W^2}{s}+1\right)\frac{1}{c_W^2}+1\right]M_W^2 C_0\left(0,0,s,M_W^2,0,M_Z^2\right) \\
& -\frac{1}{2}\left(\frac{M_Z^2}{s}+\frac{M_W^2}{s}+1\right)B_0\left(s,M_W^2,M_Z^2\right) \\
& +\left(\frac{M_Z^2}{2s}+1\right)\frac{A_0\left(M_Z^2\right)}{M_Z^2} +\left(\frac{M_W^2}{2s}+1\right)\frac{A_0\left(M_W^2\right)}{M_W^2}\bigg\}.
\end{split}
\end{align}
The counterterms for this process read:
\begin{align}
\delta_R &= 0, \\
\delta_L &= \frac{ie}{\sqrt{2}s_W} \left(\delta Z_e -\frac{\delta s_W}{s_W} +\frac{1}{2} \delta Z_W +\frac{1}{2}\left(\delta Z_{ii}^{\bar{f},L,\dagger} + \delta Z_{ii}^{f,L}\right)\right).
\end{align}
Here, the conjugated wavefunction counterterm is chosen for the
antiparticle, with the usual unchanged counterterm chosen for the
particle in the process. The tree level couplings are:%
\begin{align}
c_R = 0, \quad
c_L = \frac{ie}{\sqrt{2}s_W}.
\end{align}

\subsection{\texorpdfstring{$H\to f\bar{f}$}{H->ffbar}}

The vertex corrections to the Higgs decay into fermions are more
complex than the previously considered examples as all masses have to
be retained. The amplitude  will only be used for $H\to \ell^+\ell^-$-decays, while
the colourful decays are treated via the parton shower. 

The QED corrections to this vertex read:
\begin{align}
\begin{split}
V^{\mr{QED}} 
= \frac{\alpha}{4\pi} \frac{iem_f}{2s_WM_W}2Q_f^2s_W^2F_S^{\mr{QED}} \left(P_L+P_R\right).
\end{split}
\end{align}
The QED form factor is given by:
\begin{align}
\begin{split}
F_S^{\mr{QED}} =& \left(s-2m_f^2\right) C_0\left(m_f^2,m_f^2,s,m_f^2,0,m_f^2\right)-2B_0\left(m_f^2,m_f^2,0\right) + 1 \\
&-\frac{4m_f^2}{4m_f^2-s}\left[(B_0\left(s,m_f^2,m_f^2\right)-B_0\left(m_f^2,m_f^2,0\right)\right].
\end{split}
\end{align}
The complete weak result reads:
\begin{align}
\begin{split}
V^{\mr{weak}} = \frac{\alpha}{4\pi} \frac{iem_f}{2s_WM_W}2F_S^{\mr{weak}} \left(P_L+P_R\right).
\end{split}
\end{align}
The form factor reads:%
\begin{align}
\begin{split}
F_S^{\mr{weak}} = -M_W^2\bigg[&f_1C_0\left(m_f^2,m_f^2,s,M_W^2,m_{f'}^2,M_W^2\right) + f_2C_0\left(m_f^2,m_f^2,s,M_Z^2,m_{f}^2,M_Z^2\right) \\
&+ f_3C_0\left(m_f^2,m_f^2,s,m_{f'}^2,M_W^2,m_{f'}^2\right)+ f_4C_0\left(m_f^2,m_f^2,s,m_{f}^2,M_Z^2,m_{f}^2\right) \\
&+ h_1C_0\left(m_f^2,m_f^2,s,M_H^2,m_{f}^2,M_H^2\right) +h_2C_0\left(m_f^2,m_f^2,s,m_{f}^2,M_H^2,m_{f}^2\right)\bigg] \\
&+f_5B_0\left(s,M_W^2,M_W^2\right)+f_6B_0\left(s,M_Z^2,M_Z^2\right)+f_7B_0\left(s,m_{f'}^2,m_{f'}^2\right) \\
&+f_8B_0\left(m_f^2,M_W^2,m_{f'}^2\right)+f_9B_0\left(m_f^2,M_Z^2,m_f^2\right)+f_{10} \\
&+h_3B_0\left(s,M_H^2,M_H^2\right)+h_4B_0\left(s,m_f^2,m_f^2\right)+h_5B_0\left(m_f^2,M_H^2,m_f^2\right).
\end{split}
\end{align}
The coefficients are given by the following expressions:%
\begin{align}
f_1 =&\frac{1}{4}\bigg(\big[\big(4+w_{f'}\left(2+w_h\right)\big)\left(1-w_{f'}\right)-w_f\left(10-4w_{f'}-\left(1-2w_{f'}\right)w_h\right)\big]\mu_W^2 \\
    &+2+w_hw_{f'}-2w_f\bigg) \\
f_2 =&\frac{1}{4}\bigg(\left[\frac{4}{c_W^4}\sigma^{(2)}-w_f\left(\frac{2}{c_W^2}-w_h\right)\right]\left(\frac{1}{c_W^2}-2w_f\right)\mu_W^2 \\
  &+\frac{4}{c_W^4}v_f^2-\frac{1}{2}w_f\left(\frac{2}{c_W^2}-w_h\right)\bigg) \\
f_3 =&\frac{1}{4}w_{f'}\bigg(\big[2\left(2+w_{f'}\right)\left(1-w_{f'}\right)+2w_f\left(1+2w_{f'}-w_f\right)\big]\mu_W^2-1\bigg) \\
f_4 =&\frac{1}{4c_W^2}\bigg(\left(\sigma^{(2)}-\frac{1}{2}\right)w+w_f\left[4\sigma^{(2)}\left(\frac{1}{c_W^2}\mu_W^2+\frac{1}{2}\right)-\frac{3}{2}\right]\bigg) \\
f_5 =&\frac{1}{4}\bigg(\big[4+w_{f'}\left(2+w_h\right)-w_f\left(6-w_h\right)\big]\mu_W^2+1\bigg) \\
f_6 =&\frac{1}{4}\left(\left[\frac{4}{c_W^4}\sigma^{(2)}-w_f\left(\frac{2}{c_W^2}-w_h\right)\right]\mu_W^2+\frac{1}{2c_W^2}\right) \\
f_7 =&-\frac{1}{4}w_{f'}\bigg(2\big[2+w_{f'}-w_f\big]\mu_W^2+1\bigg) \\
f_8 =&-\frac{1}{4}\bigg(\big[2\left(2+w_{f'}\right)\left(1-w_{f'}\right)+w_{f'}w_h-w_f\left(6-2w_{f'}-w_h\right)\big]\mu_W^2+2\bigg) \\
f_9 =&-\frac{1}{4}\left(\left[\frac{4}{c_W^2}\sigma^{(2)}\left(\frac{1}{c_W^2}-w_f\right)-w_f\left(\frac{2}{c_W^2}-w_h\right)\right]\mu_W^2+\frac{2}{c_W^2}\sigma^{(2)}\right)\\
f_{10} =&\frac{1}{4c_W^2}\left(\sigma^{(2)}-\frac{1}{2}\right) \\
h_1 =&\frac{3}{2}w_fw_h\left[\left(\frac{1}{2}w_h-w_f\right)\mu_W^2-\frac{1}{4}\right] \\
h_2 =&-w_f\left[\frac{1}{8}w_h-w_f\left(w_h\mu_W^2-\frac{1}{2}\right)\right] \\
h_3 =&\frac{3}{4}w_fw_h\mu_W^2 \\
h_4 =&-w_f\left[\frac{1}{c_W^2}\sigma^{(2)}+w_f\right]\mu_W^2 \\
h_5 =&-w_f\left[\frac{3}{4}w_h-w_f\right]\mu_W^2,
\end{align}
where we used the following shorthands:
\begin{alignat}{4}
w &= -\frac{s}{M_W^2}, \quad \mu_W^2 &=& \frac{M_W^2}{4m_f^2-s}, \\
w_{f'} &= \frac{m_{f'}^2}{M_W^2}, \quad\quad  w_{f} &=& \frac{m_f^2}{M_W^2}, \\
w_h &= \frac{M_H^2}{M_W^2}. &  
\end{alignat}
Note that the corrections as written do not completely agree with
Eqs. (5.546)-(5.548) of \cite{Bardin:1999ak}. \cite{Bardin:1999ak}
provides expressions both with all masses included in
Eqs. (5.546)-(5.548), and with terms $\sim m_f^2$ neglected in
Eqs. (5.619), (5.621), (5.625). This is an appropriate approximation
for the decay $H\to b\bar{b}$ for which $m_f = m_b$, $m_{f'} = m_t$
and $m_b \ll m_t$. For our purposes, we require the exact opposite
case, with $m_f = m_\ell$, $m_{f'} = 0$. Nonetheless, the two forms can
be used to cross-check terms. In comparison to the expressions
including the full mass dependence, there is a factor of 2 in the
overall vertex in the approximated form. The latter form appears to be
correct as it reproduces the correct divergences. Secondly, the
coefficients $f_2$ and $f_6$ differ. In $f_6$, the last term should
read $\frac{1}{2c_W^2}$ instead of $\frac{2}{c_W^2}$ in agreement with
the limiting expression in Eq. (5.623). Similarly, $f_2$ has been
adapted to match the limiting expression. In particular, the second to
last term is multiplied by a factor of $\frac{4}{c_W^2}$, and the last
term by $w_f$. The second to last term can then be cast into a form
$\sim \sigma^{(2)}$ as in Eq. (5.623) by adding and subtracting
$a_f^2$. Performing these changes we find perfect agreement with \OpenLoops.

\section{Infrared form factors}
\label{sec:app:Infrared_Factors}

To complete the calculation of the infrared subtracted matrix
elements, we need the expression of the infrared factor $B_{ij}$. This
factor, and its real counterpart $\tilde{B}_{ij}(\Omega)$ are defined
in \EqRef{sec:impl:eq:B} and \EqRef{sec:impl:eq:Btilde} respectively,
have been calculated in \cite{Schonherr:2008av}, where it has also
been shown that their sum leads to a finite result as expected from
the Kinoshita-Lee-Nauenberg theorem.
In that calculation, the expressions were split up to give a number of
separate integrals which were calculated. Here, we are going to use a
different approach to also match more closely with the way the vertex
form factors are expressed. We will thus express the factor in terms
of scalar master integrals.

The virtual infrared form factor $B$ can be rewritten in the following
form, which will be more useful in expressing it in terms of master
integrals:
\begin{align}
\begin{split}
B_{ij} = -\frac{i}{8\pi^3}Z_iZ_j\theta_i\theta_j\int \mr{d}^4k \frac{1}{k^2}\Bigg[&\left(\frac{2p_i\theta_i}{k^2-2\left(k\cdot p_i\right)\theta_i} + \frac{2p_j\theta_j}{k^2+2\left(k\cdot p_j\right)\theta_j}\right)^2 \\
&- k^2\left(\frac{1}{k^2-2\left(k\cdot p_i\right)\theta_i} - \frac{1}{k^2+2\left(k\cdot p_j\right)\theta_j}\right)^2\Bigg] \label{sec:app:Infrared_Factors:eq:Bmod}.
\end{split}
\end{align}

The prefactor depends on whether the particles $i$ and $j$ are in the
initial or final state. For the purpose of this publication, we have
to consider final-final and initial-final dipole combinations. In both
cases, the factor $Z_i Z_j \theta_i \theta_j = -1$ so that the
prefactor before the integral becomes $\frac{i}{8\pi^3}$. Note that
the $p_i$ used are the momenta of the external particles. To translate
these into the momenta $q_i$ running in the loop, we use $p_1 = q_1$,
$p_2 = -q_2$ for the final state particles and $p_3 = -p_1-p_2 =
(q_2-q_1)$ for the initial state particle.

In the following we list the explicit expressions for the form factors in terms of standard
scalar integrals~\cite{Passarino:1978jh}.

\subsection{Final-Final}
\begin{align*}
B_{12} 
= -\frac{1}{4\pi}\bigg[&2\left(s-m_1^2-m_2^2\right)C_0\left(m_1^2,m_2^2,s,0,m_1^2,m_2^2\right) \\
& +2m_1^2C_0\left(m_1^2,m_1^2,0,0,m_1^2,m_1^2\right)+2m_2^2C_0\left(m_2^2,m_2^2,0,0,m_2^2,m_2^2\right) \\
& +B_0\left(s,m_1^2,m_2^2\right)-\frac{1}{2}B_0\left(0,m_1^2,m_1^2\right)-\frac{1}{2}B_0\left(0,m_2^2,m_2^2\right)\bigg].
\end{align*}
For the double virtual corrections in the decay of the $Z$-boson in
Section \ref{sec:impl:VV}, we need the infrared factor $B$ in the
limit of $s \gg m_i^2$, regulated with a small photon mass
$\lambda$. In this case, we have $m_1 = m_2 \equiv m$ and the factor
reads:
\begin{align}
\begin{split}  B 
= -\frac{\alpha}{\pi} \biggl[&-\frac{1}{2}
\log\left(\frac{\lambda^2}{m^2}\right) \log\left(\frac{-s}{m^2}\right)
+ \frac{1}{4} \log^2\left(\frac{-s}{m^2}\right) -\frac{\pi^2}{12} 
+\frac{1}{2}\log\left(\frac{\lambda^2}{m^2}\right) + \frac{1}{2}
-\frac{1}{4} \log\left(\frac{-s}{m^2}\right)\biggr]. \label{sec:app:Infrared_Factors:eq:photonmass}
\end{split}
\end{align}

\subsection{Initial-Final}
\begin{align*}
B_{31} 
= -\frac{1}{4\pi}\bigg[&2\left(s-m_2^2+m_1^2\right)C_0\left(s,m_1^2,m_2^2,0,s,m_1^2\right) 
 +2sC_0\left(s,s,0,0,s,s\right)+2m_1^2C_0\left(m_1^2,m_1^2,0,0,m_1^2,m_1^2\right) \\
& +2B_0\left(m_2^2,s,m_1^2\right)-B_0\left(0,m_1^2,m_1^2\right)-B_0\left(0,s,s\right)\bigg].
\end{align*}

\section{Real corrections}
\label{sec:app:real}

We will describe here the way we implement the real corrections. We
will describe the procedure for the decays of the vector bosons and
define the shorthand $\Gamma^{\mu} \equiv
\gamma^{\mu}\left(c_LP_L+c_RP_R\right)$, with the couplings $c_{L/R}$
given in \ref{sec:app:NLO_EW}. For the decays of a scalar boson, this
reduces instead to $\Gamma = \left(c_LP_L+c_RP_R\right)$ and we remove
the polarization vector of the decaying vector boson.

The real matrix element for the process $Z\to f\bar{f}\gamma$ reads:
\begin{align}
\begin{split}
  \mathcal{M}_1^{\frac{1}{2}} =ie^2\bar{u}\left(p_1,s_1\right) \biggr[\gamma^{\nu}\frac{\slashed{p}_1+\slashed{k}+m}{\left(p_1+k\right)^2-m^2}\Gamma^{\mu} -\Gamma^{\mu}\frac{\slashed{p}_2+\slashed{k}-m}{\left(p_2+k\right)^2-m^2}\gamma^{\nu}\biggl]v\left(p_2,s_2\right)\epsilon_{\mu}^Z(p,\lambda)\epsilon_{\nu}^{\gamma \ast}\left(k,\kappa\right).
\end{split}
\end{align}
We can express the fermion propagator as a sum over spins of an intermediate particle:
\begin{equation}
\slashed{p} \pm m = \frac{1}{2}\sum_s \left[\left(1\pm \frac{m}{\sqrt{p^2}}\right) u(p,s)\bar{u}(p,s) + \left(1\mp \frac{m}{\sqrt{p^2}}\right) v(p,s)\bar{v}(p,s)\right], \label{eq:propagatorcut}
\end{equation}
where $u[v](p,s)$ are [anti-]spinors of a fictitious fermion with mass $m = \sqrt{p^2}$. 

In the following, we will make use of a function called $X$, which is defined as:
\begin{equation}
X\left(p_1,s_1;p;p_2,s_2;c_R,c_L\right) = \bar{u}\left(p_1,s_1\right)\slashed{p} \left[c_RP_R + c_LP_L\right] u\left(p_2,s_2\right),
\end{equation}
where the $u$ may be particle or anti-particle spinors. The latter
case will be denoted through a bar over the spin index
$s_i$. Similarly, we can define another function $Y$:
\begin{equation}
Y\left(p_1,s_1;p;p_2,s_2;c_R,c_L\right) = \bar{u}\left(p_1,s_1\right) \left[c_RP_R + c_LP_L\right] u\left(p_2,s_2\right),
\end{equation}
which would be used in the decay of a Higgs boson, when there is no
structure $\Gamma^{\mu}$ in the real matrix element. The calculation
of these functions has been outlined in \cite{Krauss:2001iv} and
\cite{Schonherr:2008av}, and are based on the work in
\cite{Kleiss:1985yh}, \cite{Ballestrero:1992dv} and
\cite{Ballestrero:1994jn}.

Using these functions, we can write the full amplitude as:
\begin{align}
\begin{split}
\mathcal{M}_1^{\frac{1}{2}} =\frac{ie^2}{2} \Biggl[&\frac{1}{\left(p_{a}\right)^2-m^2}\sum_{s} \Biggl\{\left(1+\frac{m}{\sqrt{\left(p_{a}\right)^2}}\right) X\left(s_1,\epsilon^{\gamma \ast},p_{a},s\right)X\left(p_{a},s,\epsilon^{Z},\bar{s}_2\right) \\
&+\left(1-\frac{m}{\sqrt{\left(p_{a}\right)^2}}\right) X\left(s_1,\epsilon^{\gamma \ast},p_{a},\bar{s}\right)X\left(p_{a},\bar{s},\epsilon^{Z},\bar{s}_2\right)\Biggr\} \\
&-\frac{1}{\left(p_{b}\right)^2-m^2}\sum_{s} \Biggl\{\left(1-\frac{m}{\sqrt{\left(p_{b}\right)^2}}\right) X\left(s_1,\epsilon^{Z},p_{b},s\right)X\left(p_{b},s,\epsilon^{\gamma \ast},\bar{s}_2\right) \\
&+\left(1+\frac{m}{\sqrt{\left(p_{b}\right)^2}}\right) X\left(s_1,\epsilon^{Z},p_{b},\bar{s}\right)X\left(p_{b},\bar{s},\epsilon^{\gamma \ast},\bar{s}_2\right)\Biggr\}\Biggr], \label{eq:Rcut}
\end{split}
\end{align}
with
\begin{align}
p_a = p_1+k, \quad p_b = p_2+k.
\end{align}
For the double real matrix elements, to reduce the size of the
expressions, we only write the spin labels, the intermediate momenta
and the respective internal vector, so that
$X(p_i,s_i;\epsilon_j;p_k,s_k;c_{L'},c_{R'}) \equiv
X(\{p_i,\}s_i,\epsilon_j,\{p_k,\}s_k)$. For the external leptons, it
is understood that the spin label $s_i$, $i\in \{1,2\}$, corresponds
to the momentum $p_i$ and we leave the momentum out. It is further
understood, that the left- and right-handed couplings are $1,1$ when
contracted with a photon polarization and $c_L,c_R$ when contracted
with the $Z$-polarization.

For the process $Z\to f\bar{f}\gamma \gamma$, the matrix element reads:
\begin{align}
\begin{split}
  \mathcal{M}_2^1 =ie^3\bar{u}\left(p_1,s_1\right) \biggr[&\gamma^{\nu}\frac{\slashed{p}_1+\slashed{k}_1+m}{\left(p_1+k_1\right)^2-m^2}\gamma^{\rho}\frac{\slashed{p}_1+\slashed{k}_1+\slashed{k}_2+m}{\left(p_1+k_1+k_2\right)^2-m^2}\Gamma^{\mu} \\
  &-\gamma^{\nu}\frac{\slashed{p}_1+\slashed{k}_1+m}{\left(p_1+k_1\right)^2-m^2}\Gamma^{\mu}\frac{\slashed{p}_2+\slashed{k}_2-m}{\left(p_2+k_2\right)^2-m^2}\gamma^{\rho} \\
  &+\Gamma^{\mu}\frac{\slashed{p}_2+\slashed{k}_1+\slashed{k}_2-m}{\left(p_2+k_1+k_2\right)^2-m^2}\gamma^{\nu}\frac{\slashed{p}_2+\slashed{k}_2-m}{\left(p_2+k_2\right)^2-m^2}\gamma^{\rho} \\
  &+\left(k_1\leftrightarrow k_2\right)\biggr] v\left(p_2,s_2\right)\epsilon_{\mu}^Z(p,\lambda)\epsilon_{\nu}^{\gamma \ast}\left(k_1,\kappa_1\right)\epsilon_{\rho}^{\gamma \ast}\left(k_2,\kappa_2\right) .
\end{split}
\end{align}
We can repeat the procedure used for the single real matrix element,
replacing each propagator by a spin sum, ending with a large
expression which we will not reproduce here.

For the decay of a $W$ boson, we have the following real matrix element:
\begin{align}
\begin{split}
  \mathcal{M}_1^{\frac{1}{2}} =ie^2\bar{u}\left(p_1,s_1\right) \biggr[&\gamma^{\nu}\frac{\slashed{p}_1+\slashed{k}+m}{\left(p_1+k\right)^2-m^2}\Gamma^{\tau} \\
  & +\Gamma^{\mu} \frac{g_{\mu\rho}-\frac{(p-k)_{\mu}(p-k)_{\rho}}{p^2}}{(p-k)^2-p^2} V_{\tau \rho \nu}\left(p,-p+k,-k\right)\biggl]v\left(p_2,s_2\right)\epsilon_{\tau}^W(p,\lambda)\epsilon_{\nu}^{\gamma \ast}\left(k,\kappa\right),
\end{split}
\end{align}
where we introduced the triple boson vertex $V_{\tau \rho \nu} =
g_{\tau \rho}\left(p_2-p_1\right)_{\nu} + g_{\rho
  \nu}\left(p_3-p_2\right)_{\tau} + g_{\nu
  \tau}\left(p_1-p_3\right)_{\rho}$. The first term in this matrix
element can be treated like the terms in the process $Z\to
f\bar{f}\gamma$. For the second term, we first contract the triple
boson vertex, the $W$-propagator and the polarization vectors. What
remains is a structure as in the definition of the $X$-function, so we
can directly write down the result.

\bibliographystyle{JHEP}
\bibliography{bibliography}
  \end{document}